\documentclass[preprint,12pt]{elsarticle}
\usepackage{amssymb}
\usepackage{amsmath}
\usepackage{caption}
\usepackage{subcaption}
\usepackage{listings}
\usepackage{xcolor}
\usepackage{microtype}

\journal{Computer Methods and Programs in Biomedicine}

\begin{document}

\begin{frontmatter}

\title{A novel method for analysis of transient morphological changes in quasiperiodic physiological signals and their neurogenic correlates}

\author[pw]{Tomasz Gradowski} 
\ead{tomasz.gradowski@pw.edu.pl}
\author[pw,cmc]{Damian Waląg}
\author[pw,icter,ipc]{Tomir Domański}
\author[pw]{Teodor Buchner} 

\affiliation[pw]{organization={Faculty of Physics, Warsaw University of Technology},
 addressline={Koszykowa 75}, 
 city={Warsaw},
 postcode={00-662}, 
 country={Poland}}
\affiliation[cmc]{organization={Center For Digital Medicine, National Institute of Cardiology},
 addressline={Alpejska 42}, 
 city={Warsaw},
 postcode={04-628}, 
 country={Poland}}
\affiliation[icter]{organization={International Centre for Translational Eye Research},
 addressline={Skierniewicka 10A},
 city={Warsaw},
 postcode={01-230},
 country={Poland}}
\affiliation[ipc]{organization={Institute of Physical Chemistry, Polish Academy of Sciences},
 addressline={Kasprzaka 44/52},
 city={Warsaw},
 postcode={01-224},
 country={Poland}}

\begin{abstract}

\subsection*{Background and Objective}
Transient changes in electrocardiographic (ECG) morphology often accompany fluctuations in heart rhythm and may provide clinically relevant information about cardiac function. 
However, conventional ECG visualization and analysis methods typically emphasize either waveform morphology or rhythm variability, making it difficult to assess their temporal interplay in long-term recordings and, consequently, to identify rhythm-related morphological changes and exploit their potential diagnostic value. This work presents a visualization framework for quasiperiodic physiological signals that enables simultaneous assessment of beat-to-beat morphological changes and rhythm dynamics in a single representation.

\subsection*{Methods}
The proposed method converts quasiperiodic signals into two-dimensional carpet plots. Characteristic events (e.g., ECG R peaks) are used to align consecutive signal segments, which are transformed into color-coded rows and stacked in chronological order. The resulting image preserves both the temporal evolution of signal morphology and variations in cycle duration. The method was evaluated using ECG recordings from multiple publicly available databases containing healthy subjects and patients with diverse cardiac abnormalities, as well as synchronized multimodal physiological recordings.

\subsection*{Results}
Carpet plots enabled rapid visualization of transient morphological changes alongside heart rate dynamics across recordings ranging from several minutes to hours. The representation highlighted clinically relevant phenomena, including ST segment alterations, QT interval variability, changes in T wave morphology, atrial fibrillation episodes, premature ventricular complexes, Wenckebach periodicity, and stress-test phase transitions. The image-based representation was also shown to be suitable for automated analysis using convolutional neural network feature extraction.

\subsection*{Conclusions}
Carpet plots provide a compact representation of quasiperiodic physiological signals, jointly visualizing rhythm and morphology across long-term recordings. The proposed framework facilitates both expert interpretation and image-based computational analysis, offering a general approach for investigating transient physiological phenomena in ECG and other synchronized quasiperiodic signals.

\end{abstract}

\begin{keyword}
ECG \sep quasiperiodic signals \sep signal processing \sep signal visualization \sep electrophysiology \sep ST segment analysis \sep QT analysis \sep morphology analysis \sep beat-by-beat analysis

\end{keyword}

\end{frontmatter}


\section{Introduction}
\label{sec:intro}

In signals from many areas of science and technology, quasi-repeating patterns appear, with subsequent forms differing slightly. We may represent this phenomenon as the result of two interacting processes. The first of them determines the time instants at which the pattern appears. Formally speaking, this is a point process, also used in physiology \cite{Turcott1994}, a type of stochastic process that defines intervals on the time axis; we refer to it as a rhythmic process. The second process defines the specific pattern and its morphology; we refer to it as the morphological process. Many rhythmic processes can be represented this way: the electrocardiogram (ECG), the course of blood pressure, plethysmography, or intracranial pressure and other cardiogenic rhythms in human and animal organisms, the activity of a neuron or their group, patterns in active centers in lasers, forest fires, and other phenomena occurring in active media, often described by spatiotemporal reaction-diffusion equations. 

In the case of ECG, the rhythmic process is the well-known heart rate variability (HRV) \cite{taskforce1996hrv}, which is predominantly governed by the autonomic nervous system (ANS) and to a much lesser extent by the internal dynamics of the cardiac effector\footnote{Cardiac conductance time affects the PR interval, and hence contributes to the RR interval dynamics - this effect is most often of minor importance.}. The sinus node, the heart's natural pacemaker, initiates the rhythmic process. It is subject to constant neural and hormonal control \cite{Boyett2000}, which modulates its chronotropic action \cite{Brubaker2011}. This point process is fully described by the sequence of time intervals between successive events in a single realization. Peripheral blood pressure follows the cardiac cycle and is also affected by pulse transit time \cite{Hoshide2022-hw}.

A morphological process is a description of a single evolution. In the case of the ECG, it represents the electrical activity of the heart muscle. It is a complex spatiotemporal biological and physicochemical process based on ionic displacements, carried out by the cardiac conduction system and cardiac muscle tissue in response to stimulation. It may be observed via potential imaging \cite{Efimov,WILDOWICZ2025199}, surface electrocardiogram or magnetocardiogram (MCG) \cite{Peczalski2024}, or invasively \cite{Kuklik2009}.
Likewise, in the realm of electronics, there exists a class of monostable multivibrators, which, when triggered, generate a specific pattern, such as an exponential drop to zero. Thus, any dynamics containing quasiperiodic patterns is controlled by two coupled processes: one with continuous time and one with discrete time.

\subsection{Clinical importance of joint rate-morphology analysis.}
\label{subsec:clinical}

It is interesting to analyze the clinical context using cardiac morphology and rate analysis as examples. It is often necessary to understand the coupling between rhythmic and morphological processes, namely, how momentary changes in timing or frequency translate into morphological alterations. 

In ECG, an example of transient behavior is ST segment elevation or depression \cite{Wagner2009}. Such an analysis typically focuses on a single feature, such as the amplitude of the maximum ST departure from baseline (elevation or depression). Experienced clinicians, however, know that this is the bottom line of morphological analysis.
The ST segment elevation is important, but many concomitant changes may accompany this departure: e.g., the wandering of the J point \cite{Junttila2012}, widening of the S wave \cite{DeAndrade2022}, or even the amplitude of the R wave \cite{Andrzejewska2022}, which also have clinical significance. Typically, a trend of a single feature is tracked \cite{Bjerregaard2003}. The ability to monitor many trends across different features is technically limited by the visualization method. Correlation with heart rate changes is typically visualized as a 2-D scatterplot.

Morphological changes may alternate from beat to beat. ST-segment depression or elevation may be accompanied by other beat-to-beat morphological changes \cite{Kessler1992}. T wave alternans affects every other beat (period doubling) and originates from underlying changes in action potential duration (APD) \cite{Glass1991}. Alternans is a rare event of period doubling with profound consequences for the electrical stability of heart function \cite{Narayan2006, Krogh-Madsen2007, Starobin2009, Cutler2009, Aro2016, Gabr2023}.

In many cases, it is sufficient to look at the signal; however, there also exist certain general-purpose techniques, both in the time and frequency domains, that are also applied to the ECG morphology assessment, including but not limited to Principal Component Analysis (PCA) \cite{Okin2002}, Singular Value Decomposition (SVD) \cite{Andersen2007}, Wavelet Transform \cite{GualsaquiMiranda2016}, and Fast Fourier Transform (FFT) \cite{Maniewski1996}.

Numerous visualization methods have previously been proposed for ECG and other physiological signals. Conventional ECG displays preserve waveform morphology but become impractical for long recordings. Tachograms and Poincaré plots \cite{Li2011} effectively visualize rhythm dynamics but intentionally discard most morphological information. Time-frequency representations, including spectrograms, scalograms \cite{PENG2004199} and wavelet maps \cite{Stiles2004}, emphasize frequency content rather than beat-aligned morphology. Recurrence plots \cite{Mewett1999} characterize nonlinear dynamics but do not preserve the morphology of individual cardiac cycles.

However, changes in morphology may be associated with either simple or more complex beat-by-beat dynamics.
Heart rate dynamics alone can exhibit substantial complexity. Numerous analytical techniques for linear and nonlinear dynamics have been developed to characterize rhythmic processes \cite{taskforce1996hrv, Acharya2006}. These approaches, however, are limited to rhythm analysis, focusing on interval variability while disregarding the signal's morphological features. There are rare exceptions to this rule in the context of nonlinear analysis \cite{Baranowski2002, Baumert2015}. 

Accurate QT assessment itself remains challenging, particularly when QT prolongation is heterogeneous, and recent work has explored AI-based approaches to improve automated QT measurement \cite{Kim2026}. A particularly challenging example of a strong relationship between rhythm and morphology, QT-RR dynamics, can be studied using an RR-QT diagram \cite{Malik2002}. Such a diagram does not allow us to observe changes over time, and we lose the ability to see whether the trajectory clusters around a moving point attractor or instead jumps between distant points. If we consider the cellular-level counterpart of QT-RR dynamics, the APD/RR or APD/DI analysis provides important information about the stability of electrical activity \cite{Hastings2000}. It is essential to assess the interplay between the rate and the morphology. 

Recent studies further demonstrate the growing potential of AI-based ECG analysis for detecting clinically relevant abnormalities, including acute coronary occlusion, highlighting the value of representations that preserve rich ECG information for automated analysis \cite{SHARKEY2026102671}.

The main contribution of this work is not merely a new visualization of ECG signals, but a general analytical framework for representing quasiperiodic physiological signals in a form that jointly preserves rhythmic and morphological information. Existing approaches typically emphasize either temporal dynamics, such as heart-rate variability analysis, or waveform morphology, such as beat-by-beat ECG inspection. The carpet plot integrates these two complementary aspects into a single, synchronized representation, enabling direct observation of their interaction across recordings ranging from several minutes to 24-hour of Holter monitoring.

The specific contributions of this work are fourfold:
\begin{enumerate}
    \item A general method for transforming quasiperiodic signals into synchronized two-dimensional representations that applies beyond ECG to other physiological signals.
    \item A unified representation enabling simultaneous assessment of rhythm and morphology, facilitating the analysis of transient coupling between these processes.
    \item A visualization framework suitable for long-term recordings, allowing rapid identification of transient and evolving physiological phenomena.
    \item An image-based representation naturally compatible with modern computer vision and machine learning techniques for automated analysis.
\end{enumerate}

\section{Method}

To address the limitations of existing approaches, we propose a general framework for representing quasiperiodic signals in a synchronized two-dimensional form. Rather than analyzing rhythmic variability and waveform morphology separately, this method aligns individual cycles according to characteristic trigger events and visualizes their evolution over time within a common coordinate system. This representation preserves both the temporal organization of successive cycles and their detailed morphology, enabling direct analysis of interactions between these two components. This approach enables detailed tracking of how the features of quasi-repeating patterns evolve, capturing both gradual and abrupt morphological changes. By aligning individual pattern realizations according to their rhythmic triggers, the method facilitates comparison of successive cycles, allowing researchers to observe intra- and inter-cycle variability with high temporal resolution. Additionally, it reveals dependencies among parameters, such as amplitude, duration, and waveform shape, within and across cycles. Crucially, the method enables quantitative assessment of how variations in the frequency or timing of pattern appearances (i.e., fluctuations in the rhythmic process) influence the signal's morphology, thereby revealing underlying coupling mechanisms.

We demonstrate the utility of this method using ECG signals, which exhibit clear quasiperiodic structure governed by the cardiac rhythm and rich morphological variability across heartbeats. By applying the analysis framework, individual QRS complexes are extracted and aligned by R peak timing, enabling precise visualization of morphological dynamics over time. This alignment reveals subtle changes in waveform features such as QRS duration, ST segment elevation, and T wave morphology, features that may indicate physiological adaptations or early signs of pathology. Furthermore, by correlating instantaneous heart rate (derived from RR intervals) with changes in the shape and amplitude of ECG components, we can quantify the coupling between rhythm variability and morphological responses. This application highlights the method's potential to enhance diagnostic resolution and monitoring in clinical settings, particularly for arrhythmia detection, stress testing, and assessment of autonomic function. It is particularly useful for identifying transient changes caused by chemical imbalance, which may occur when a high pacing rate does not allow the morphological process sufficient time to maintain homeostasis. Note that the coupling is bidirectional: sometimes morphological changes precede rhythm changes; typically, they follow.

It should be emphasized that, to maintain generality, we do not propose any specific diagnostic marker for an electrophysiological symptom of a particular medical condition. Instead, we demonstrate the approach's generality, with the expectation that further application of the method will enable the identification of disease-specific phenomena through exploratory analysis.

A limited version of the method has previously been applied to ECG analysis: either in the context of AF diagnosis (ECHOView \cite{Krasteva2025}) or in the general ECG context, limited to two consecutive beats (Electrocardiomatrix \cite{Li2015}). We will analyze all the differences between the methods in the Discussion section.

\begin{figure}
     \centering
     \includegraphics[width=\textwidth]{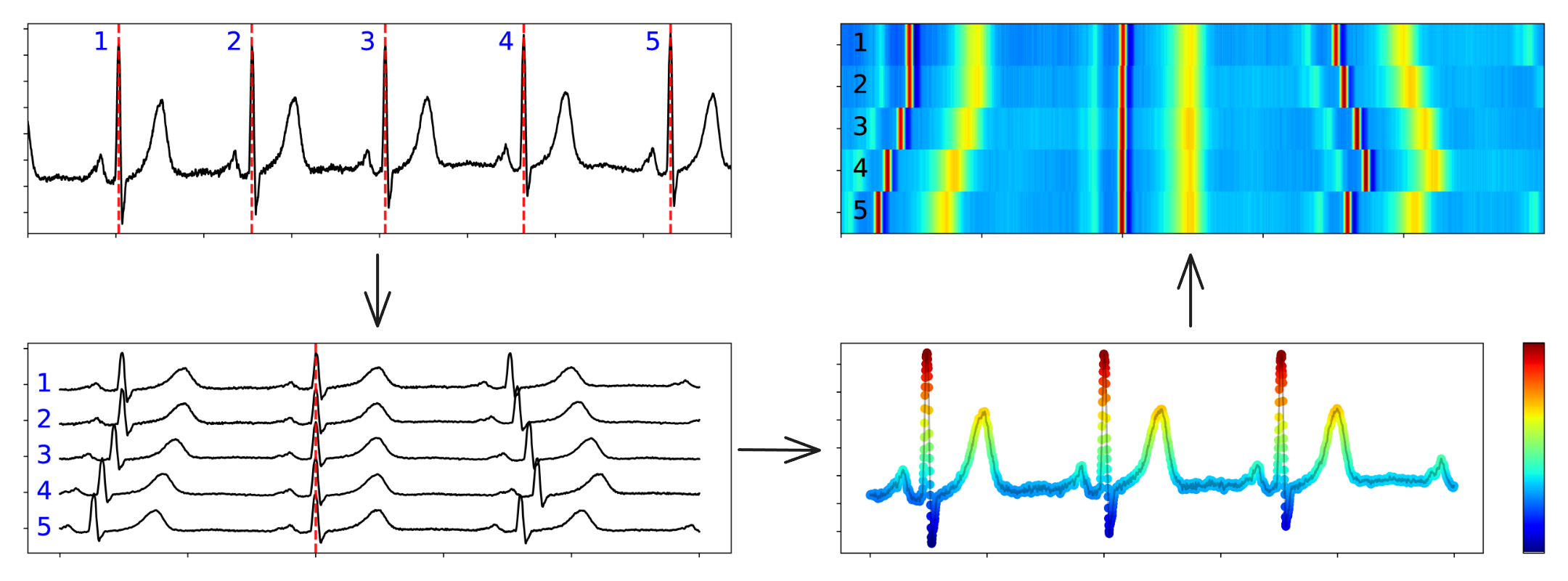}
    \caption{Flowchart of the carpet-plot generation procedure.
Original ECG signal with marked R peaks $\rightarrow$ fixed-duration segments aligned by the central R peak $\rightarrow$ amplitude-to-color mapping applied within each segment $\rightarrow$ final carpet plot formed by stacking segments in chronological order.}
        \label{fig:method}
\end{figure}

The proposed method transforms quasiperiodic signals into a graphical representation. The main stages of the procedure are summarized in Fig. \ref{fig:method}. 

To construct the visualization of a signal, the following procedure is applied:

1. Selection of a characteristic feature:
The first step is to identify a characteristic feature of the quasiperiodic pattern, typically one that shows high correlation across different realizations.

2. Segmentation of the signal:
The signal is segmented into temporal windows based on the identified characteristic points. Each segment is centered around a specific occurrence of the pattern -- referred to as the central pattern -- and includes either the preceding or following pattern, or both. These segments have a fixed duration, ensuring that the central pattern appears at the same relative position within each segment. This consistency is critical for enabling meaningful comparisons between cycles.

3. Amplitude-to-color mapping:
Within each segment, the amplitude value of every sample is mapped to a corresponding color using a predefined transfer function. This mapping can be either linear or nonlinear, depending on the desired sensitivity and contrast. The result is a color representation of the waveform, where subtle amplitude differences become visually distinguishable.

4. Construction of the visualization (carpet plot):
Finally, all individual segments are stacked vertically in chronological order, aligned by the central pattern's position, forming a two-dimensional image we refer to as a carpet plot. In this image, each horizontal line corresponds to a single cycle segment, and the colors along it encode the segment's amplitude. The vertical axis represents the progression of cycles over time, while the horizontal axis corresponds to the relative time within each segment.

The above method may be applied to ECG signals, using the R peak as the characteristic feature to identify and align quasiperiodic patterns. For each detected R peak, a segment of the ECG signal is extracted, spanning a window from e.g., 1 second before to 1.5 seconds after the peak -- capturing the full morphological context of the cardiac cycle. This results in segments of uniform duration (2.5 seconds in this case), with the R peak consistently positioned within each segment. Each sample within a segment is then mapped to a color. 

Color mapping is performed in two stages. First, an amplitude range to be represented by the color scale is selected. This range may be defined by the signal's minimum and maximum values, but in practice, we generally prefer percentile-based clipping to reduce the influence of isolated artifacts. Importantly, the selected amplitude limits are determined for the signal (or selected analysis range) and are not recalculated independently for each cardiac cycle. Thus, the same amplitude-to-color mapping is applied to all rows of a given carpet plot, allowing changes in color between consecutive rows to reflect actual changes in signal amplitude rather than beat-specific normalization.

The selected amplitude interval is then mapped to a chosen color scale. In this study, we primarily used the \texttt{jet} colormap from the Matplotlib library \cite{Hunter2007} for its broad color range and high visual contrast. However, the choice of colormap is not intrinsic to the proposed method and may be adapted to the analytical purpose. For example, grayscale mappings may be preferable for printed figures, whereas a diverging scale such as blue-white-red (\texttt{bwr}) can be useful when deviations from a defined reference level are of interest, with the reference value assigned to the center of the scale. Similarly, the amplitude limits can be deliberately restricted to a specific morphological component. For example, selecting limits appropriate to the T wave amplitude range and clipping values outside this interval increases the color resolution available for visualizing T wave morphology.

The segments are subsequently arranged vertically in time order to form a carpet plot, where each horizontal line corresponds to a single heartbeat and its surrounding context. The resulting visualization enables rapid assessment of how ECG morphology evolves, with carpet plot durations ranging from a few minutes (e.g., 3 minutes) to long-term recordings (e.g., 24 hours).

Importantly, the RR intervals are not normalized in the proposed representation. No interpolation, time-warping, or zero-padding is applied to make consecutive cardiac cycles equal in duration. Instead, a fixed-duration temporal window is extracted around each detected R peak, while the original time scale within this window is preserved. Consequently, the position of the subsequent R peak varies from row to row according to the actual RR interval. This preserved RR variability is an essential part of the representation, not an artifact to be removed; it allows rhythm-related changes in morphology, including changes in QT duration and ST segment morphology, to be interpreted together with the corresponding changes in cardiac cycle length.

The following Python code snippet illustrates the implementation of the carpet plot generation process:
   
\begin{lstlisting}[language=Python]
def make_carpet(ecg, rpeaks, preR, postR):
 return [color_scale(ecg[r-preR:r+postR]) 
 for r in rpeaks]
\end{lstlisting}

In this code, \texttt{ecg} is the input ECG signal, \texttt{rpeaks} is a list of indices corresponding to detected R peaks, and \texttt{preR} and \texttt{postR} define the number of samples before and after each R peak to include in the segments. The function constructs a two-dimensional array in which each row corresponds to a segment centered on an R peak, enabling subsequent visualization. Note that the boundary checks in the above code snippet are omitted for clarity.

While the selected parameters used in this study provide a practical and effective configuration, they are not fixed and can be adjusted to suit specific analytical goals or clinical requirements. In particular, the pre-R interval can be shortened if needed. However, it is crucial to ensure the P wave is fully captured, as it contains diagnostically significant information about atrial activity. The post-R interval may be extended in exceptional cases, such as in patients with severe bradycardia, where cardiac cycles are prolonged. Additionally, while the \texttt{jet} colormap offers high visual contrast, grayscale colormaps may be preferable in certain contexts, especially when figures are to be printed in grayscale, as they maintain readability and preserve essential morphological distinctions.

Another important point to address is the baseline problem. It is known that one of the typical sources of noise in ECG recordings is baseline wandering, which may result from variable relative impedance between the electrodes that constitute the ECG lead. If this noise is not removed during signal processing, baseline correction can be performed, e.g., by setting the mean value of the PQ segment to zero, as typically done in Holter recordings. Then, naturally, this value becomes a setpoint for the whole color scale. Other options are also possible, e.g., setting the mean value of each segment to zero.

To further enhance the readability of carpet plots, especially when working with long-term ECG recordings such as Holter monitor data or stress test signals, which are often affected by artifacts, it is helpful to apply amplitude clipping during color mapping. This involves limiting the range of signal values mapped to colors, effectively suppressing extreme outliers that can distort the visual representation. A practical approach is to clip amplitude values based on percentiles; for example, the 1st--99th percentile range can be calculated from the entire signal represented by the carpet plot. The resulting limits are then kept fixed for all cardiac cycles in that carpet. Values below the lower limit and above the upper limit are assigned the colors corresponding to the respective limits. This technique preserves the visibility of typical morphological features while minimizing the influence of transient noise or artifacts, making subtle patterns easier to discern across hundreds or thousands of cycles. It is beneficial when comparing ECG morphology over extended periods, where maintaining consistent contrast is crucial for practical interpretation.

In addition, important considerations arise when carpet plots are used not only for visual inspection but also as inputs to automated analysis pipelines, such as machine learning or pattern recognition algorithms. In such cases, it is essential to maintain a constant number of rows across all carpet plots. This means that a fixed number of segments must be selected. When working with datasets that include signals recorded at different sampling rates, it is equally important to standardize the number of columns in the carpet plots -- that is, the number of samples per segment. Since each segment corresponds to a fixed temporal window (e.g., 2.5 seconds), variations in sampling frequency will result in different segment lengths unless corrected. To ensure consistency, all signals must be resampled to a common target sampling rate before visualization. This preprocessing step ensures that each segment contains the same number of data points and, therefore, that each carpet plot has a uniform width (i.e., the number of columns). Ensuring uniform input dimensions is a fundamental requirement for most machine learning models, particularly convolutional neural networks (CNNs), which expect input tensors of consistent shape.

In order to demonstrate the merits of the method, we have applied it to
selected recordings from THEW database E-HOL-03-0480-013 \cite{Couderc2010}, MIT-BIH Atrial Fibrillation Database \cite{Moody1983}, INCART database \cite{Goldberger2000}, EPHNOGRAM database \cite{Kazemnejad2021} and 
CHARIS database \cite{Kim2016}. 
The carpet plots were generated using custom-built software that enables analysis of signals in various formats \cite{github}.

\section{Results}

\begin{figure}
\centering

\begin{subfigure}[b]{0.49\textwidth}
\centering
\includegraphics[width=\textwidth]{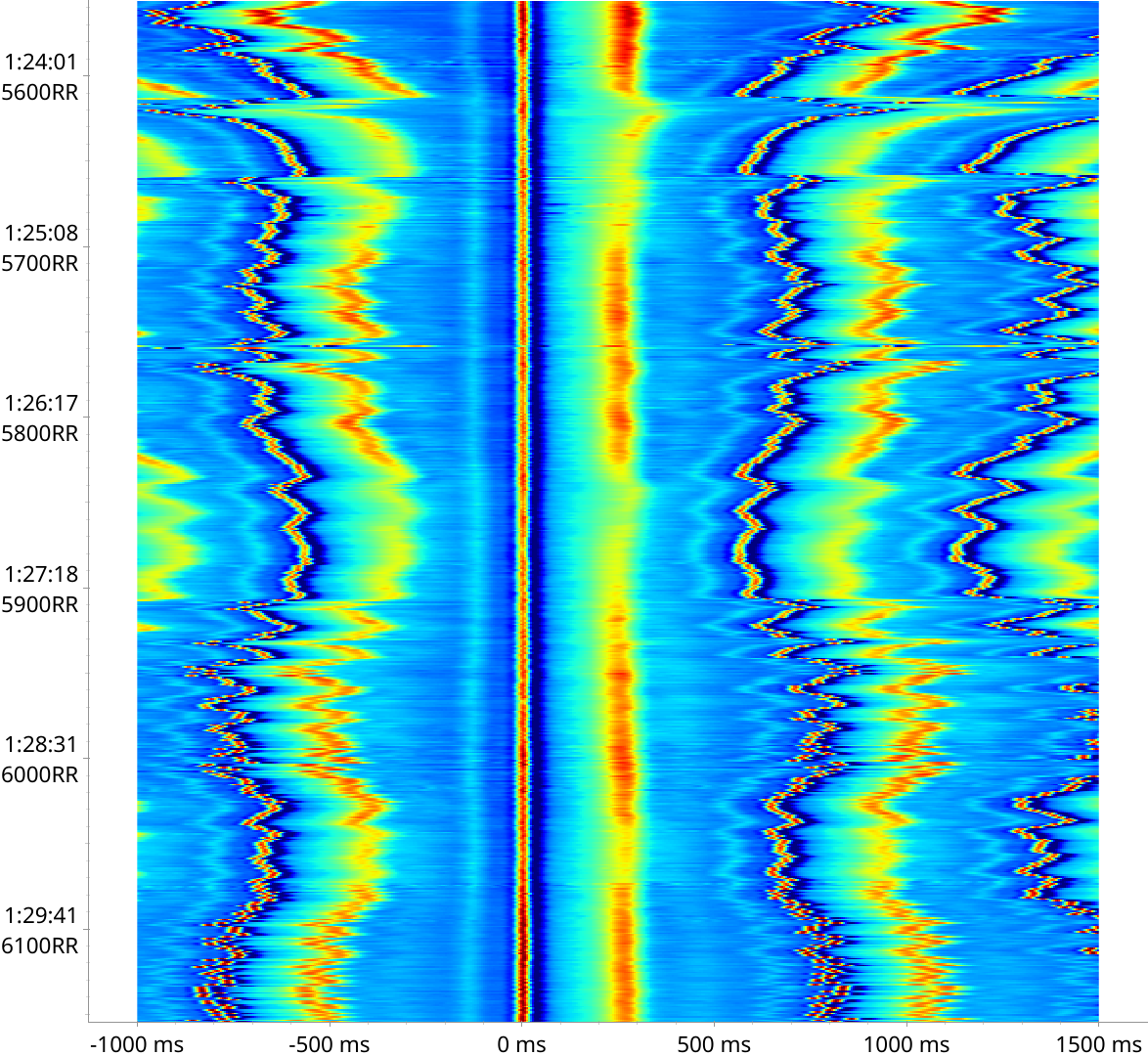}
\caption{}
\label{fig:carp1}
\end{subfigure}
\hfill
\begin{subfigure}[b]{0.49\textwidth}
         \centering
\includegraphics[width=\textwidth]{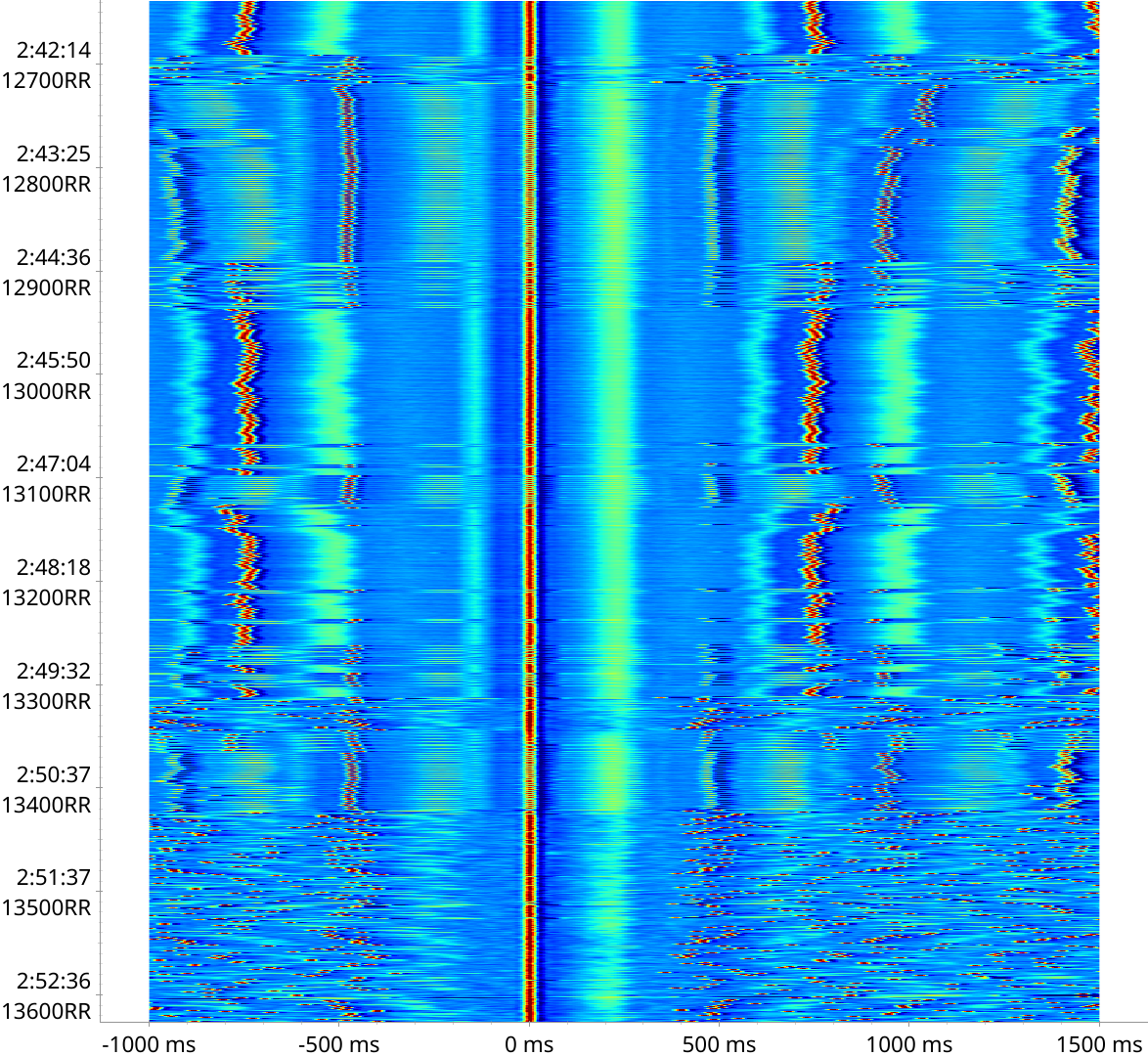}
\caption{}
\label{fig:carp2}
\end{subfigure}
\hfill
\begin{subfigure}[b]{0.49\textwidth}
         \centering
\includegraphics[width=\textwidth]{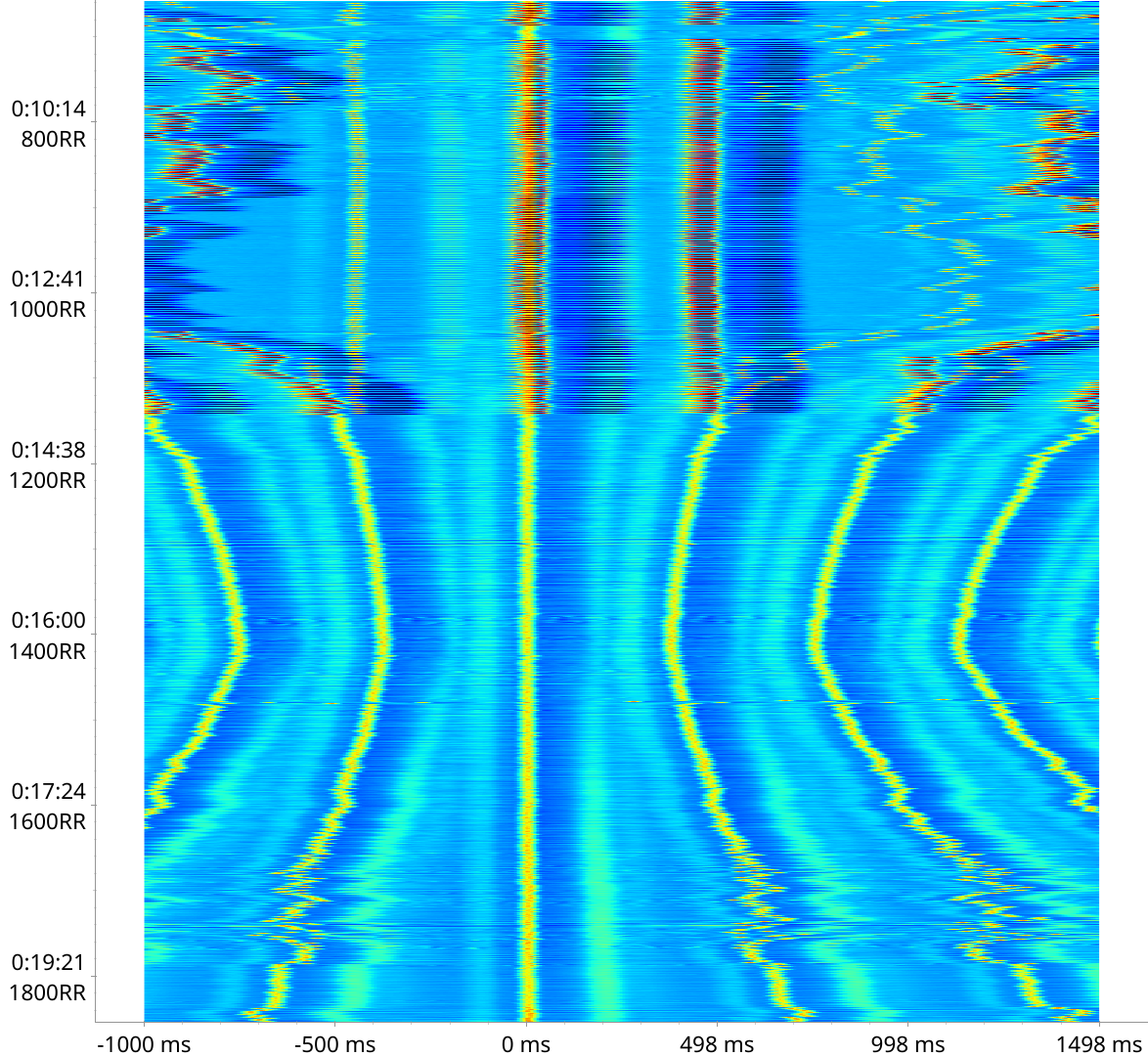}
\caption{}
\label{fig:carp3}
\end{subfigure}
\hfill
\begin{subfigure}[b]{0.49\textwidth}
         \centering
\includegraphics[width=\textwidth]{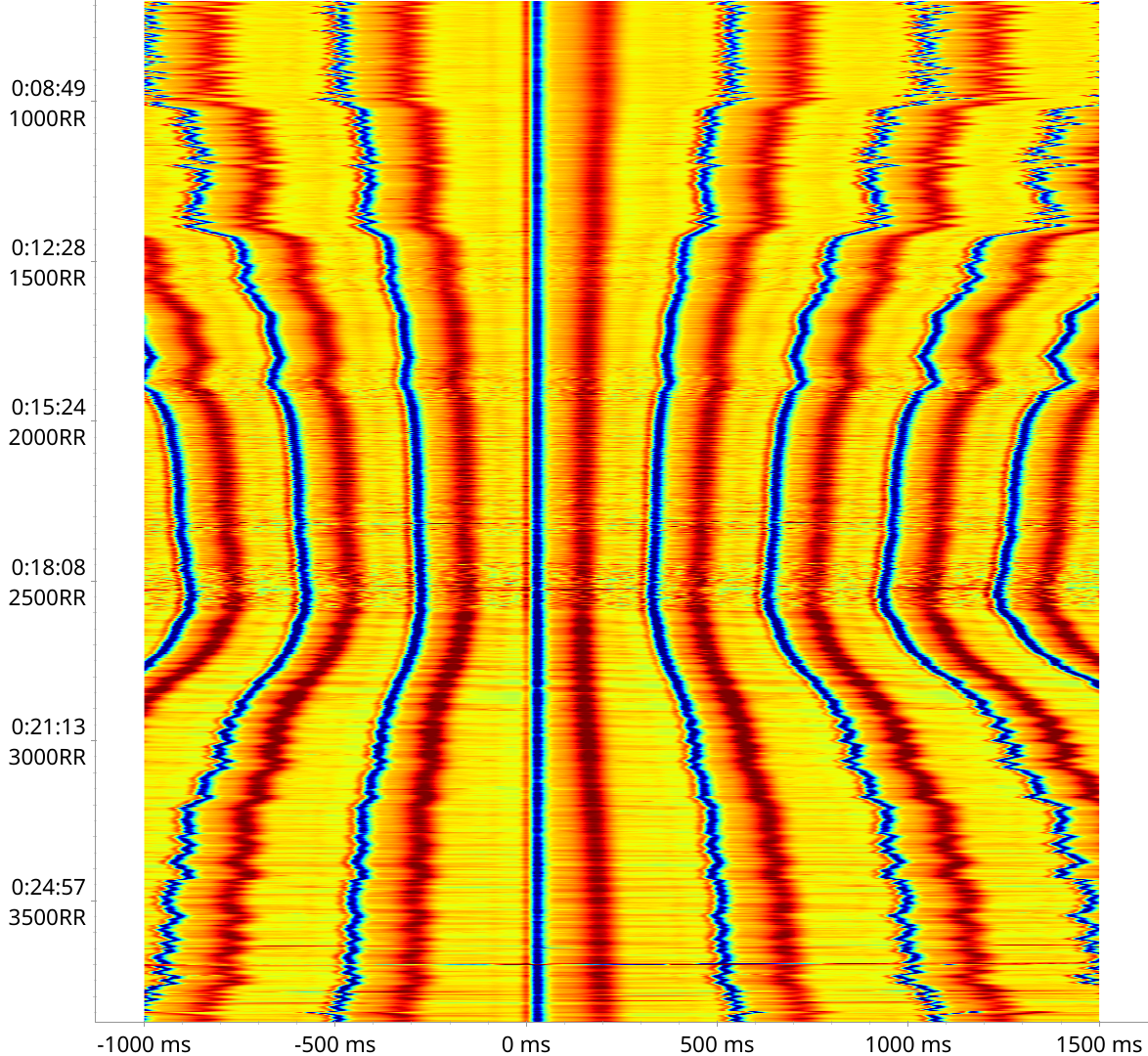}
\caption{}
\label{fig:carp4}
\end{subfigure}

\caption{Examples of carpet plots for different ECG signals. (a) male, 19 years old, diagnosed with LQT1 (patient \#174 from THEW database E-HOL-03-0480-013 \cite{Couderc2010}); (b) patient \#08405 from MIT-BIH Atrial Fibrillation Database \cite{Moody1983}; (c) record I51 (lead II) from INCART database \cite{Goldberger2000}; (d) record \#36 of stress-test session from EPHNOGRAM database \cite{Kazemnejad2021}}
\hfill
\label{fig:carpets}
\end{figure}

We evaluated the visualization method using ECG recordings from multiple publicly available databases. The chosen datasets varied in both recording length and sampling rate, demonstrating the method's robustness across heterogeneous acquisition settings. Notably, the examples included patients with diverse cardiac conditions, such as inherited arrhythmia syndromes, atrial fibrillation, and exercise-induced changes, highlighting the versatility of carpet plots in capturing both rhythm irregularities and morphological dynamics. The results are presented in Fig. \ref{fig:carpets}.

\begin{figure}
\centering
\includegraphics[width=0.49\textwidth]{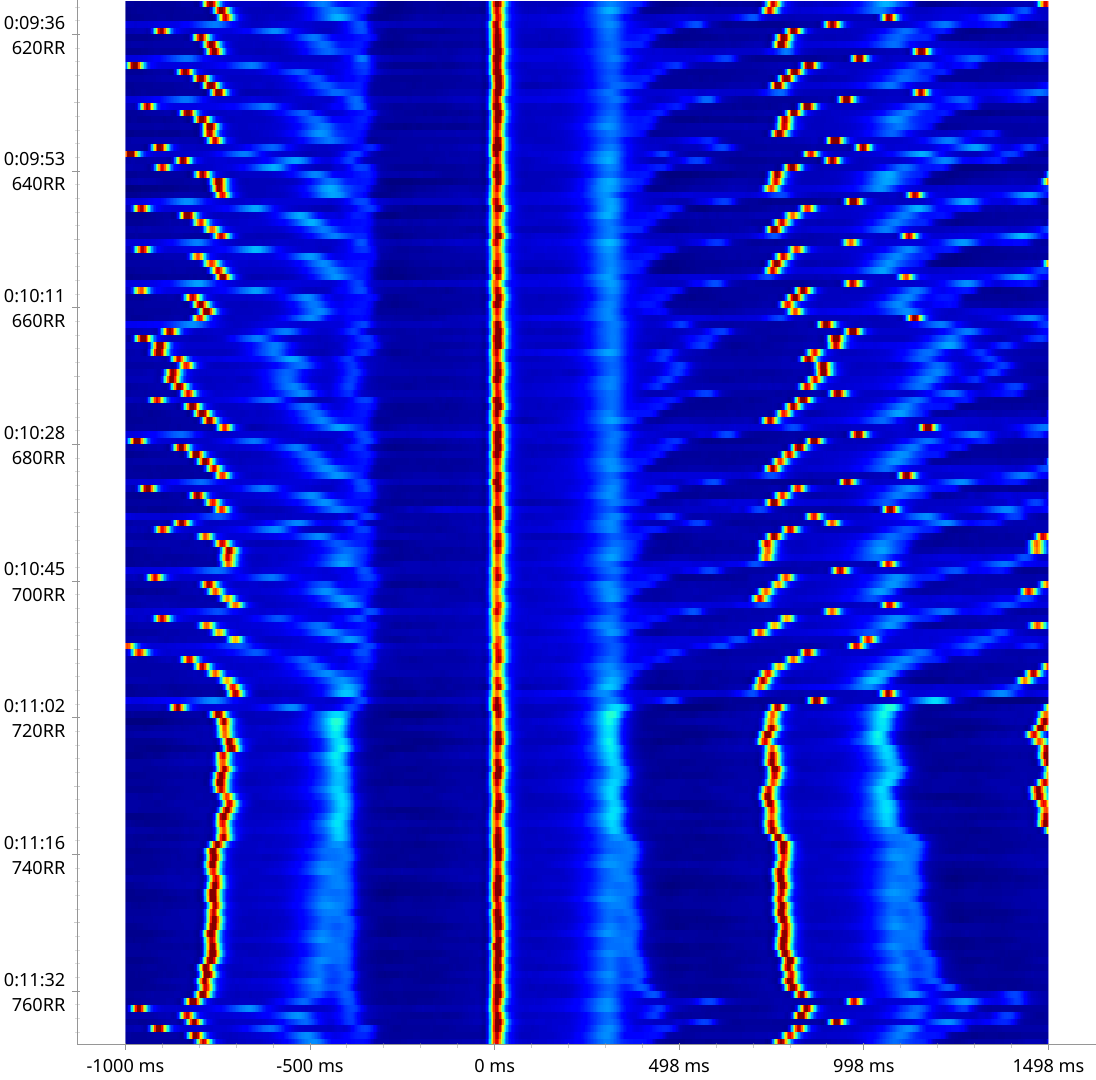}
\includegraphics[width=0.49\textwidth]{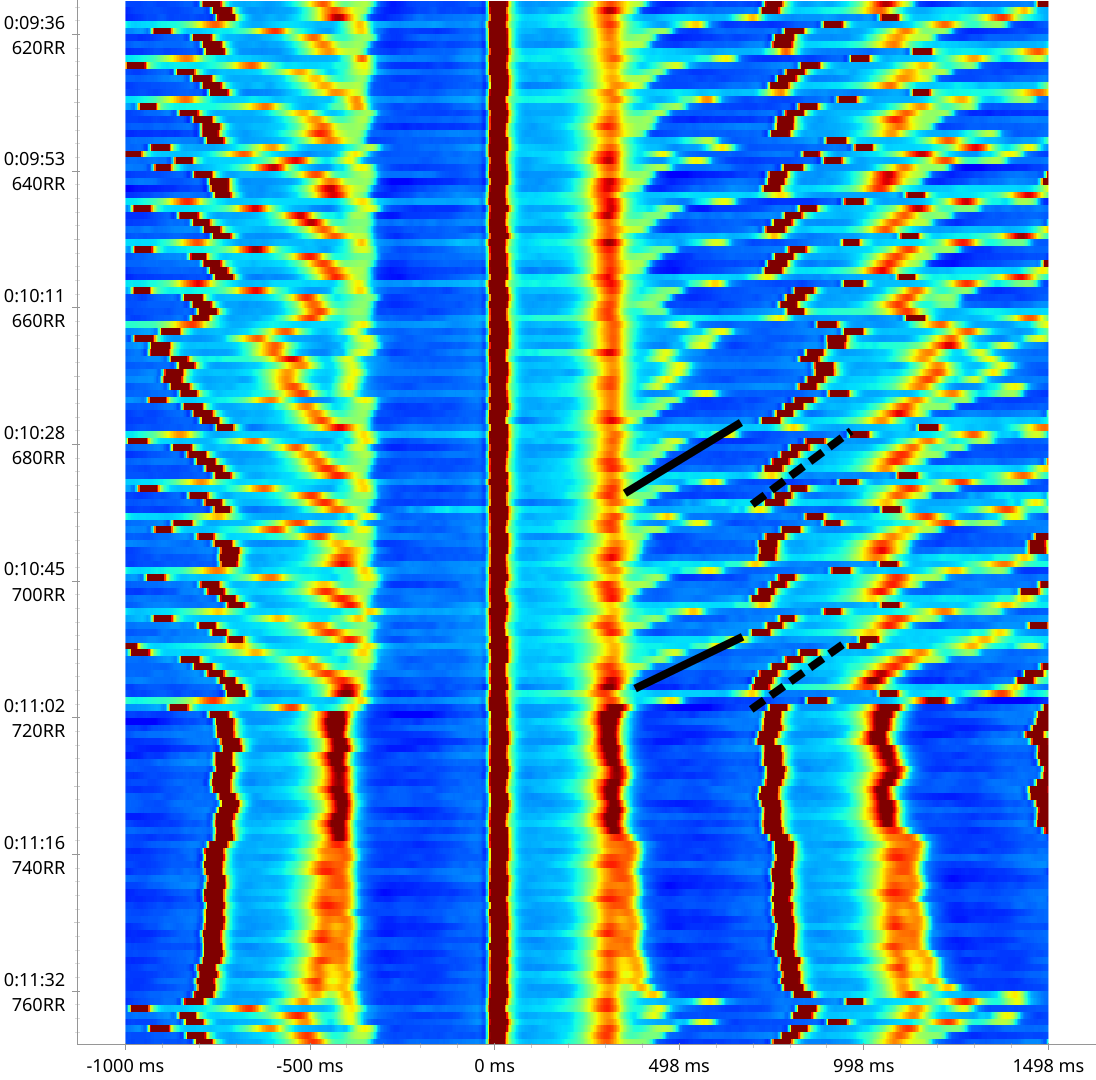}
\includegraphics[width=0.99\textwidth]{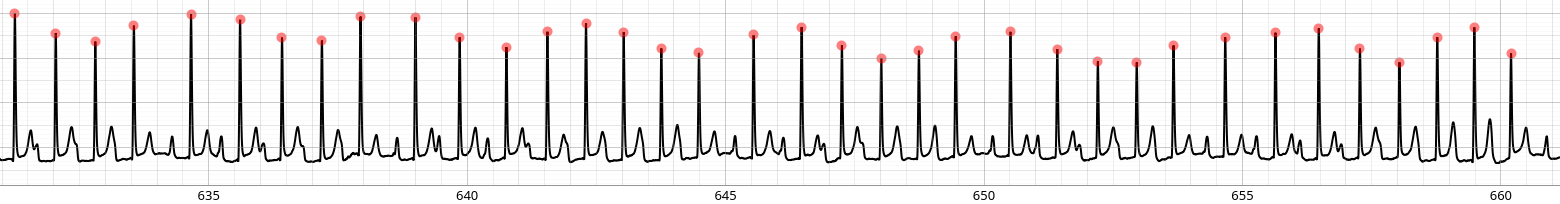}

\caption{Carpet plots of a patient diagnosed with second-degree AV nodal block (record I11 from the INCART database). Comparison of two different color mappings using the same ECG signal: with a top and bottom percentile cutoff (left) and with manually selected amplitude limits focusing on the range typical of the T and P waves (right). In both cases, the selected amplitude-to-color mapping is applied consistently across all beats. The manually restricted range increases the color resolution for lower-amplitude waveform components, making the Wenckebach periodicity clearly visible.}
\label{fig:colormapping}
\end{figure}

To demonstrate the impact of color mapping as a research tool, we present an example of a second-degree AV block with Wenckebach periodicity in the standard and adjusted color scales (Fig. \ref{fig:colormapping}). The rate of change of the PQ interval can be assessed by direct measurement of the angle of the oblique line. As conduction through the AV node becomes increasingly impaired, the rate of change in mean HR (also visible in the carpet view) provides a convenient measure of the imbalance, enabling quantification of this pathology. Hence, color-scale adjustments allow focusing on specific dynamical features.

Since the method is novel, special attention must be paid to interpreting carpet plots. The vertical axis represents successive cardiac cycles, aligned with both absolute time (HH:MM:SS) and RR interval indices. This dual labeling supports interpretation across short (minutes) and long (hours or even 24-hour) recordings. For example, Fig. \ref{fig:carpets} shows signals ranging in length from about 7 minutes (Fig. \ref{fig:carp1}) to about 25 minutes (Fig. \ref{fig:carp4}), while Fig. \ref{fig:apb_icp} shows a recording fragment lasting about 2 hours. This method also allows for the quick assessment of 24-hour signals without losing any information from the original signal. Because RR intervals vary, the vertical scale in real time is, in fact, nonlinear. Each row is anchored at the R position, and the difference between them is the RR interval, which is highly variable. The horizontal axis represents relative time within each cycle and is displayed on a linear scale. In our study, we used a 2.5-second window (1 second before and 1.5 seconds after the R wave), adjusted slightly depending on sampling rate. Heart rate patterns (and the corresponding RR intervals with their dynamics) are easily inferred by examining the position of the next R wave relative to the central R wave. This enables rapid estimation of tachograms directly from the carpet. When rotated 90 degrees counterclockwise, the carpets resemble conventional heart rate variability plots (tachograms), facilitating interpretation.

The carpet plot of the LQTS patient (Fig. \ref{fig:carp1}) demonstrates gradual and abrupt changes in T wave morphology together with QT interval variability. The widening and narrowing of the T wave region (yellow-to-red structure at around 250 ms) illustrate transient morphological changes that are difficult to appreciate when browsing long ECG recordings in the conventional format.

The next recording (Fig. \ref{fig:carp2}) illustrates two distinct rhythm transitions within approximately 15 minutes. During the first 8 minutes, the patient alternates repeatedly between sinus rhythm and ventricular bigeminy, producing characteristic periodic changes in the RR interval pattern while preserving the underlying ECG morphology. At approximately 2:51 (recording timestamp), the rhythm changes abruptly to an episode of atrial fibrillation, which appears as a sudden loss of the regular beat-to-beat organization and markedly increased RR irregularity.

The recording in Fig. \ref{fig:carp3} demonstrates ventricular ectopy occurring in bigeminal and trigeminal patterns. The premature ventricular contractions form characteristic repeating motifs that are immediately recognizable across many consecutive beats (top part of the plot). At the same time, the transition to supraventricular tachycardia is visible as an abrupt change in both rhythm and morphology.

The stress-test recording (Fig. \ref{fig:carp4}) clearly separates resting, exercise, and recovery phases. Progressive shortening of RR intervals, accompanied by gradual changes in QT duration and T wave morphology, illustrates how the carpet plot simultaneously visualizes autonomic adaptation and morphological response.

\begin{figure}
\centering
\includegraphics[width=0.49\textwidth]{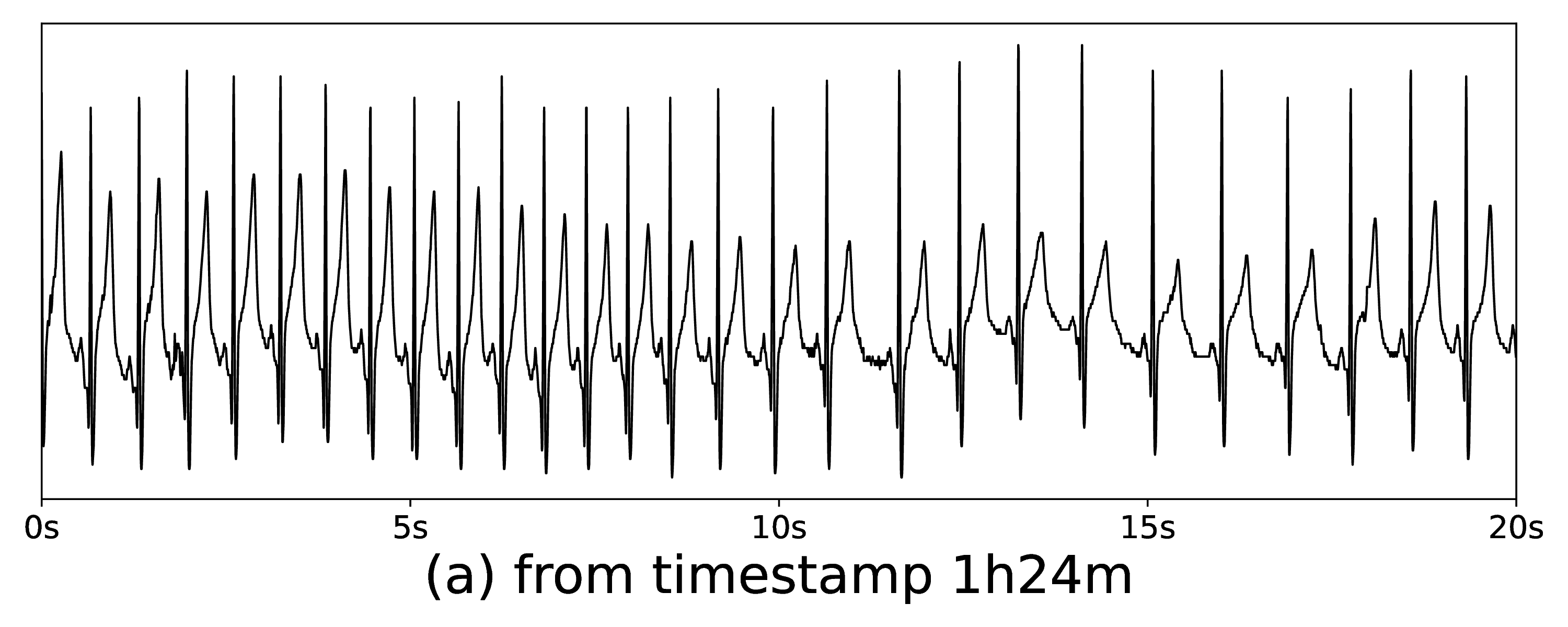}
\includegraphics[width=0.49\textwidth]{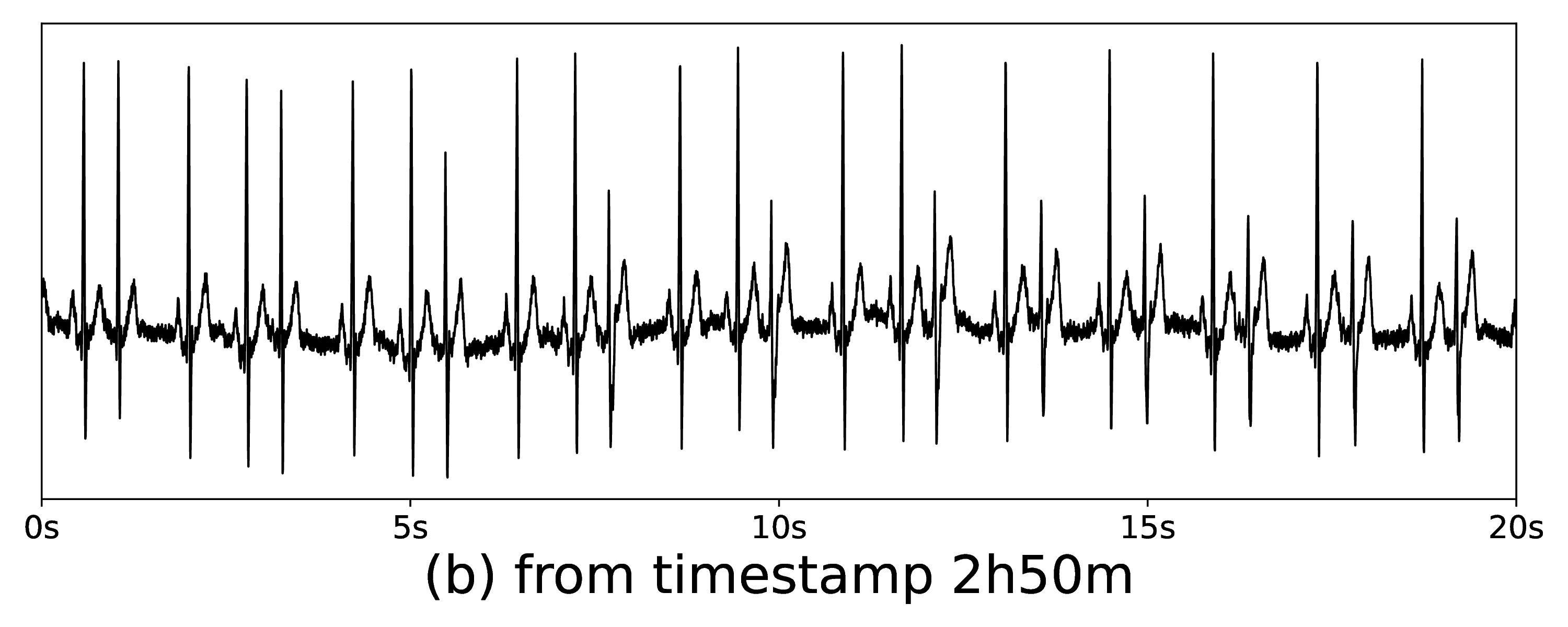}
\includegraphics[width=0.49\textwidth]{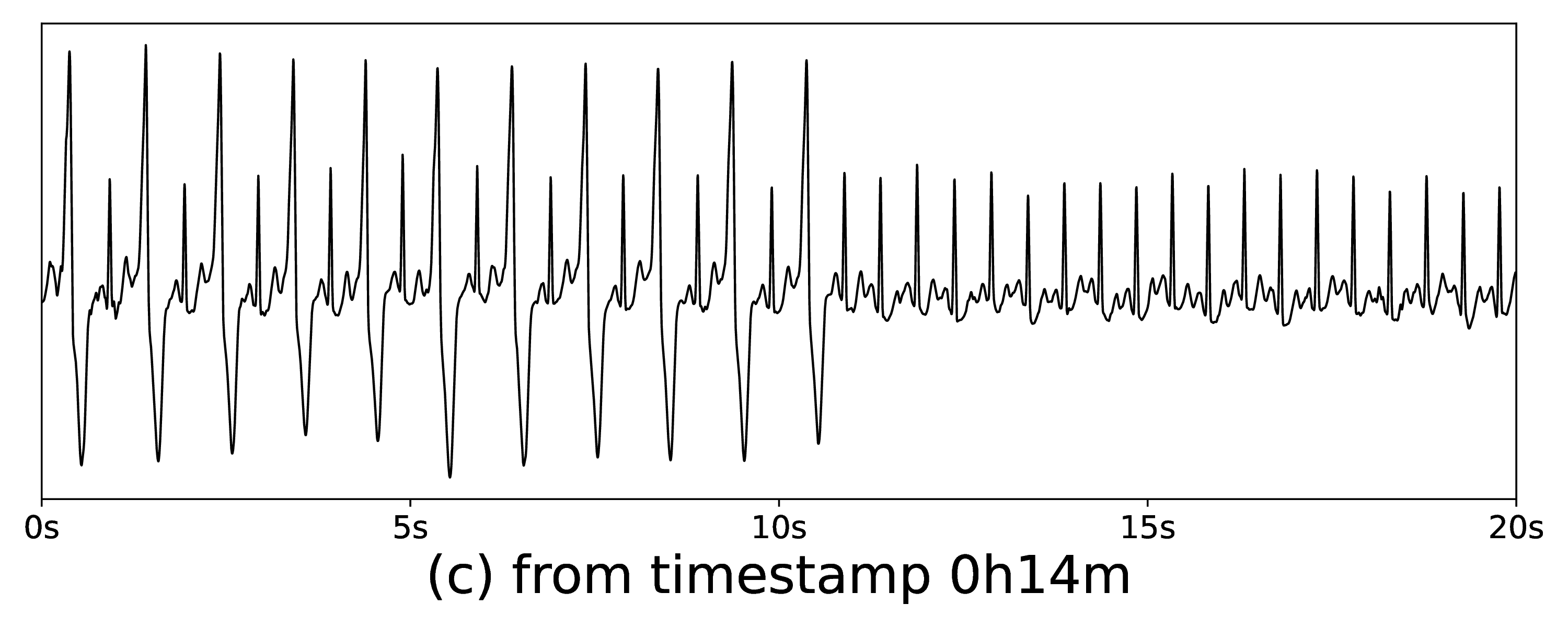}
\includegraphics[width=0.49\textwidth]{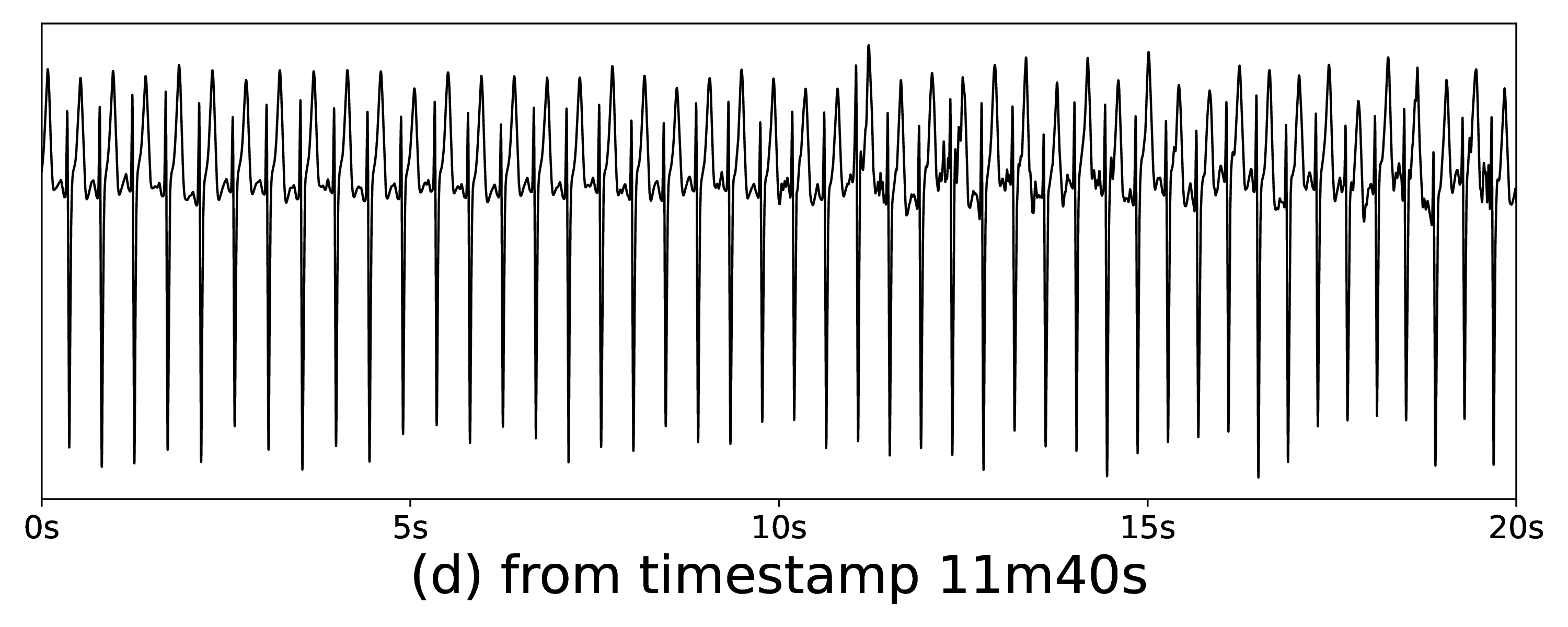}
\caption{Representative 20-second raw ECG signal segments corresponding to the carpet plots in Fig. \ref{fig:carpets}.}
\label{fig:raw_signals}
\end{figure}

To facilitate interpretation, Fig. \ref{fig:raw_signals} shows the corresponding 20-second raw ECG fragments for each dataset presented in Fig. \ref{fig:carpets}.
These short signal excerpts highlight characteristic morphological patterns -- such as QRS shape, T wave amplitude, and baseline variations -- that, when visualized over longer durations, give rise to the textures and color patterns observed in the carpet plots. By comparing Figures \ref{fig:carpets} and \ref{fig:raw_signals}, one can directly relate the features of the original waveform to their manifestation in the two-dimensional representation.

\begin{figure}
\centering
\includegraphics[width=0.99\textwidth]{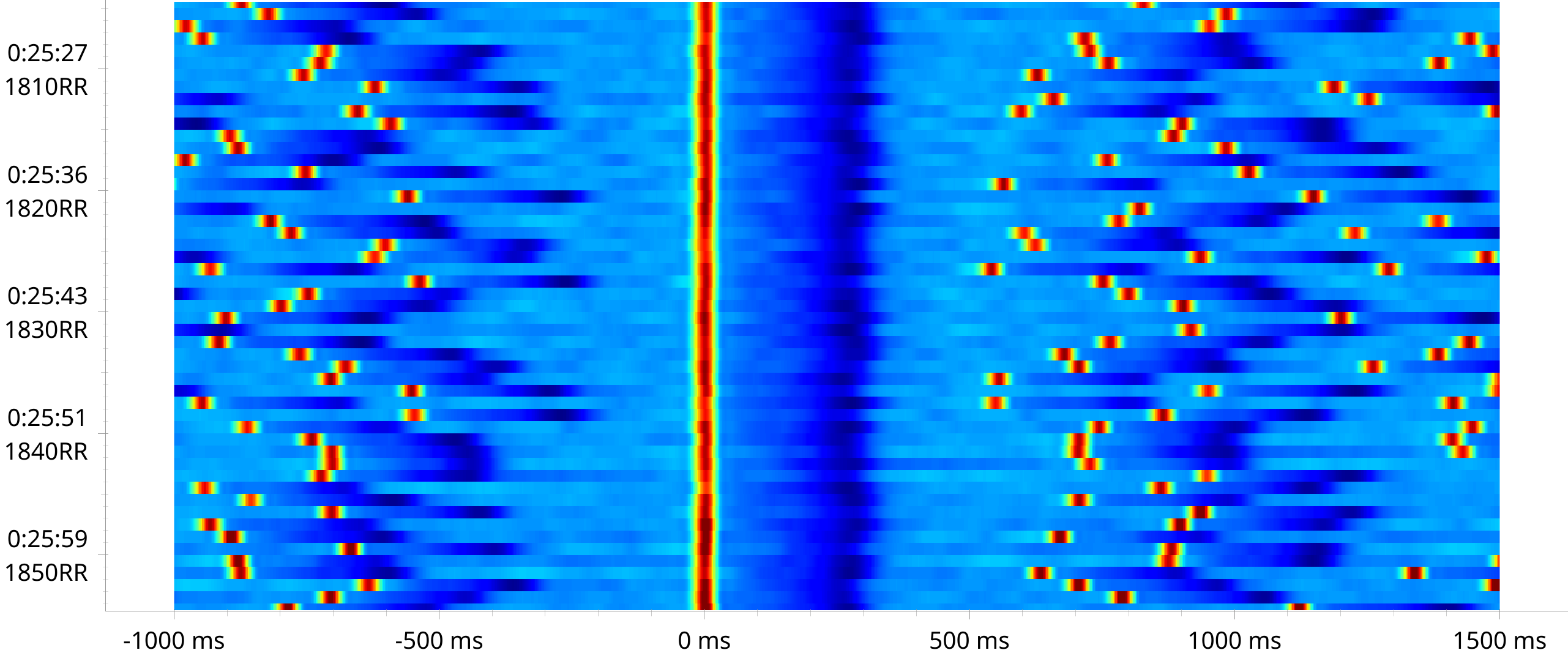}
\includegraphics[width=0.99\textwidth]{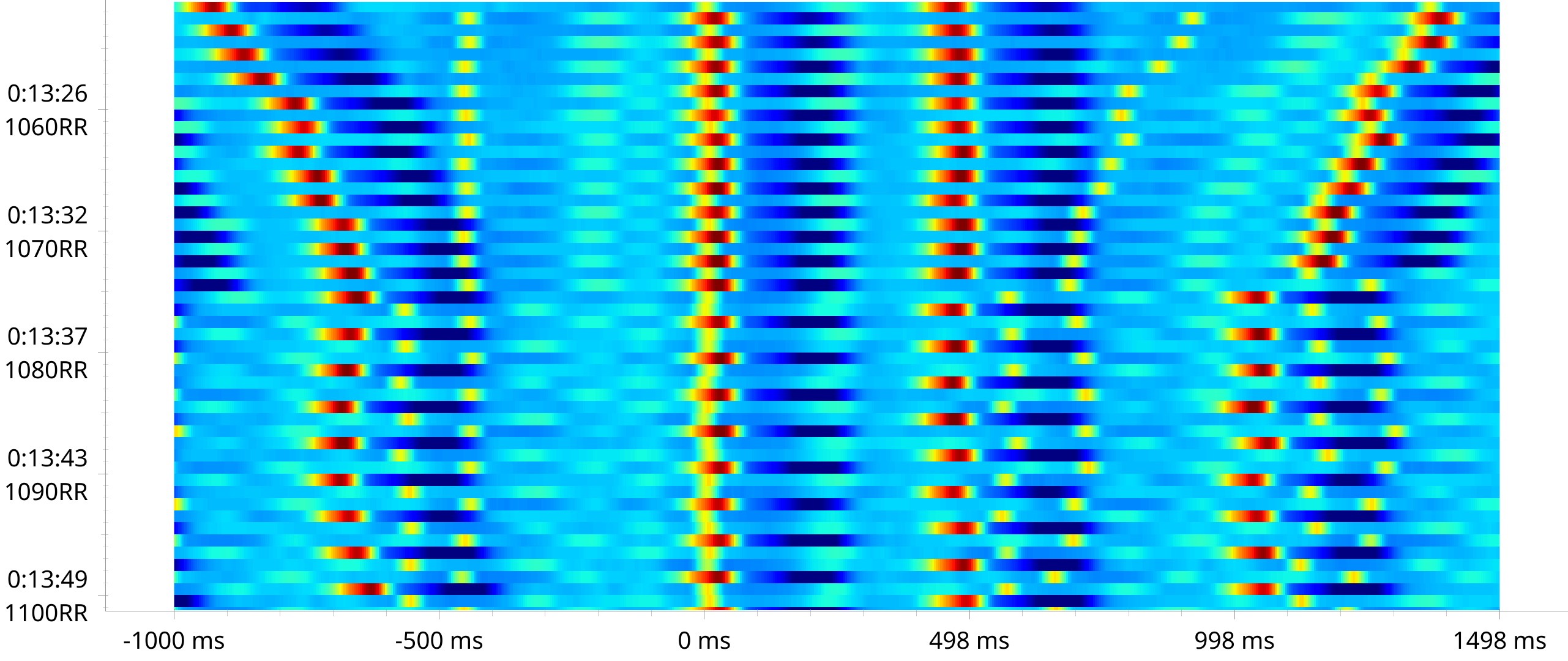}
\includegraphics[width=0.99\textwidth]{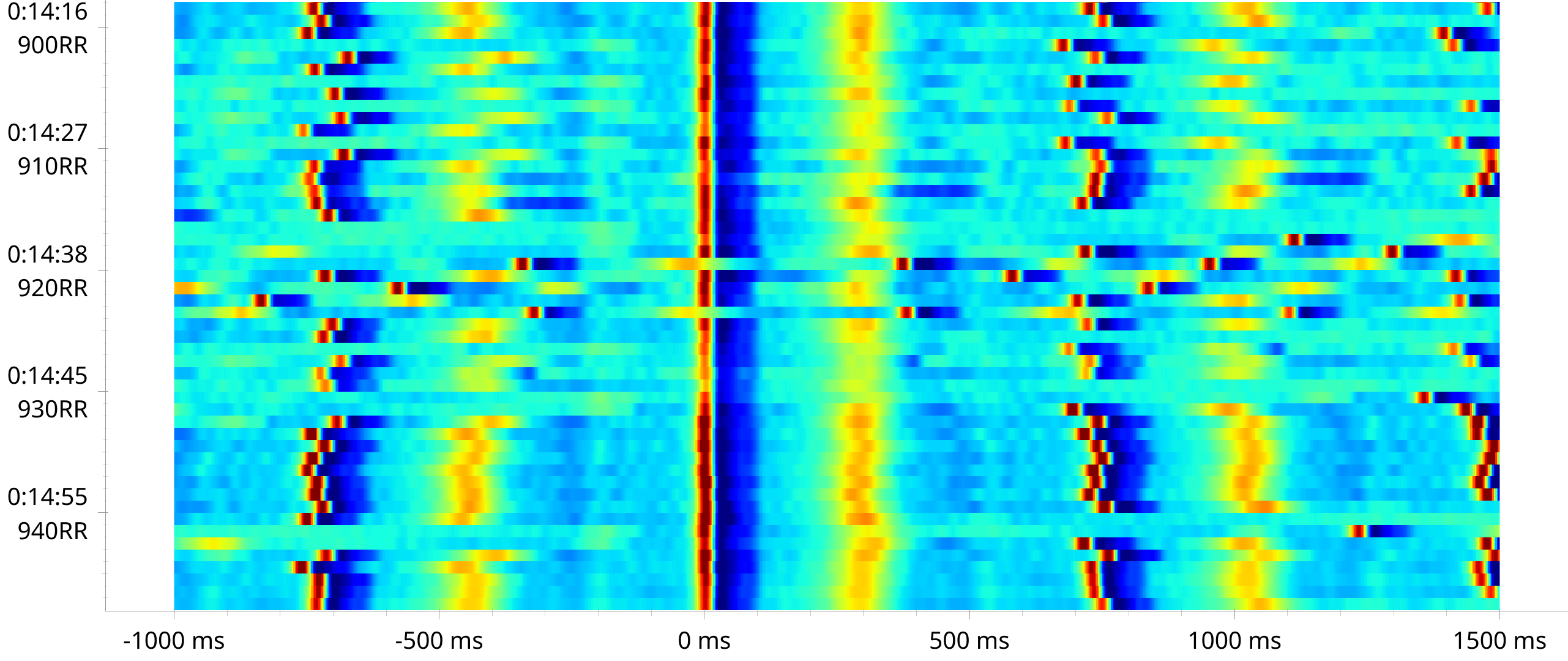}

\caption{Carpet plots (50 beats each) showing different phenomena: (top) atrial fibrillation (record 219 from MIT-BIH Atrial Fibrillation Database); (middle) bigeminy and trigeminy (record I51 from INCART database); (bottom) a long pause (record 232 from MIT-BIH Arrhythmia Database).}
\label{fig:zoom}
\end{figure}

Figure \ref{fig:zoom} presents characteristic patterns observed in several common rhythm abnormalities. Each arrhythmia produces a distinctive visual structure that can be recognized directly from the carpet representation. In atrial fibrillation (Fig. \ref{fig:zoom}, top), the highly irregular RR intervals generate a disordered arrangement of successive beats, reflecting the absence of organized atrial activity. Ventricular bigeminy and trigeminy (Fig. \ref{fig:zoom}, middle) produce a regular alternation between sinus and premature ventricular beats, giving rise to a repetitive interlace-like pattern that is immediately recognizable. Finally, a prolonged pause (Fig. \ref{fig:zoom}, bottom) appears as an abrupt interruption of the otherwise regular rhythm (only one beat per line), followed by resumption of the normal beat sequence. These examples demonstrate that carpet plots preserve both rhythm and morphology, enabling different cardiac abnormalities to be distinguished by their characteristic visual patterns.

Morphological changes are visible as vertical structures, most prominently the QRS complexes and subsequent T waves. Color coding enables the detection of subtle variations in amplitude, duration, or waveform shape. For instance, in long QT syndrome (Fig. \ref{fig:carp1}), carpets reveal abrupt changes in T wave morphology and QT interval length, which correlate with heart rate variations. The same patient also exhibits changes in the PQ interval, manifested as variations in the width of the blue band between the P wave and the QRS.

All these changes are also visible in the ECG. However, the carpet plot is, in our opinion, much better suited to the reality of an automatic ML-based event detection pipeline: note that all the findings presented in Fig. \ref{fig:carpets} were selected from large datasets by visual inspection, which was fast and effective, concentrated on departures from a typical pattern, i.e., the anomalies.

Carpet plots also reveal phase transitions in rhythm and morphology. Fig. \ref{fig:carp4} shows a recording from a stress test, where the subsequent phases of the test are clearly visible: rest phase, stress (of varying intensity), and recovery. Within each of these phases, distinct differences are visible in the rhythm (slower and more strongly correlated with breathing at rest, faster but more stable during stress) and morphology (e.g., shortening of the QT interval during stress). Phase transitions can also be abrupt (discontinuous). For example, the recording in Fig. \ref{fig:carp2} shows a patient who suddenly experiences an episode of atrial fibrillation. Fig. \ref{fig:carp3} shows a recording in which PVCs suddenly disappear when an episode of supraventricular tachycardia occurs.

Other, more subtle phenomena observed in the carpets include ST segment depression and elevation. These are manifested by a change in color or intensity on either side of the T wave (see Fig. \ref{fig:carp1}). This allows for precise observation of episodes of ischemia and the changes related not only to the elevation or depression of the ST segment, but also to transient changes in QRS complex morphology and the level of the J point, which typically pose a challenge for ST algorithms performing automated analysis.

Note that, typically, in the second-degree block, we consider the variable PQ segment within a constant RR interval; however, due to altered AV conduction time, both the PQ and RR intervals are prolonged, as a rising PQ increases RR. This effect is marked by oblique line segments: solid for the consecutive P wave peaks and dashed for the R peaks during the intermittent block episode. This is a subtle electrophysiological effect, typically not considered, as in standard projection, it is barely visible. Even in a standard color map, it can easily be left unattended. Additionally, the R peak at 0 ms exhibits an island-like structure, reflecting strong respiratory modulation of the R amplitude, as shown in the ECG.

\section{Discussion}

The principal novelty of the proposed framework lies in combining two complementary descriptions of quasiperiodic physiological signals that have traditionally been studied independently: rhythmic dynamics and waveform morphology. Rhythm analysis characterizes the temporal organization of successive cycles, whereas morphology analysis focuses on the detailed shape of individual cycles. Existing visualization techniques generally emphasize one of these domains at the expense of the other. Carpet plots preserve both simultaneously by aligning successive cycles using characteristic trigger events and displaying their complete morphology in chronological order. This enables direct observation of interactions between rhythm and morphology that are difficult to recognize using conventional visualization techniques.

In the cardiological context, previous methods have focused on rhythm characteristics, including HRV and the statistical properties of RR intervals, as well as potential rhythm disturbances, thereby providing insights into autonomic regulation and cardiovascular dynamics. The presented approach, in contrast, focuses on morphological features of individual cardiac cycles -- such as the shape, amplitude, and duration of QRS complexes, T waves, or ST segments (which are often analyzed to detect structural or electrophysiological abnormalities) -- as well as their temporal evolution and relationship with heart rate.

In the standard 12-lead projection, morphology analysis can be applied to a single evolution or to multiple evolutions. In contrast, rhythm analysis typically requires a longer strip of a single selected electrode, typically lead II.
Concerning the transient changes of certain features, they are typically analyzed as time series: one series per feature.
Moreover, the choice of features is limited, as there is typically no good reason to consider, e.g., the elevation or depression of the J point.
As a result, this type of analysis is not included in the equipment used in clinics, even for 12-lead Holter recordings.

In a scientific context, we can study as many features as we want. If the morphology is rich, as in the ECG examples presented, the number of time series required to capture the complex changes grows rapidly, making it more cumbersome to identify correlations among the observed features. By enabling simultaneous, temporally aligned analysis of both rhythm and morphology, our method bridges these two domains. The structure of the carpet plot, which aligns morphological data to rhythmically defined anchor points (i.e., R peaks), enables direct visualization and quantification of correlations between cycle-to-cycle timing variability and morphological changes. This opens up new possibilities for studying the coupling mechanisms between both processes, the rhythmic and the morphological, and for reasoning about causality, for example, whether variations in heart rate induce changes in waveform shape, or vice versa. Such joint analysis provides a richer, more holistic understanding of physiological dynamics, particularly in complex conditions where both timing and shape are affected. 

Another significant strength of our method is its suitability for both qualitative and quantitative analysis. From a qualitative perspective, the carpet plot provides a highly intuitive and condensed representation of even very long-term physiological signals, making it significantly more accessible for human interpretation than traditional raw signal plots. This is particularly valuable in scenarios involving extended recordings, such as 24-hour Holter ECGs or multi-phase stress tests, where visual analysis of the raw waveform is impractical. A single glance at a carpet plot can reveal high-level patterns, such as transitions between rest and activity, changes in heart rate variability, or periods of stable versus irregular morphology (e.g., episodes of paroxysmal atrial fibrillation). Sleep stages, physical exertion, or transient episodes of autonomic imbalance become immediately apparent, offering clinicians rapid insight without exhaustive signal scrolling or reliance on any single-purpose algorithm. The above-mentioned features make the carpet plot a convenient tool for exploratory analysis when it is not known in advance which heart rate changes and their morphological footprints will be elicited.

Simultaneously, the same carpet plot structure naturally lends itself to quantitative analysis using computational methods. Since the plots are two-dimensional, image-like representations with uniform structure, they are especially well-suited for processing by modern machine learning techniques, including convolutional neural networks (CNNs), vision transformers (ViT), and other pattern recognition models. Automated feature-detection algorithms may identify additional changes and their correlates that could be difficult to recognize through manual inspection alone. This approach has been recently utilized in the AF detection context by Krasteva \textit{et al.} \cite{Krasteva2025}.

\begin{figure}
\centering
\includegraphics[width=\textwidth]{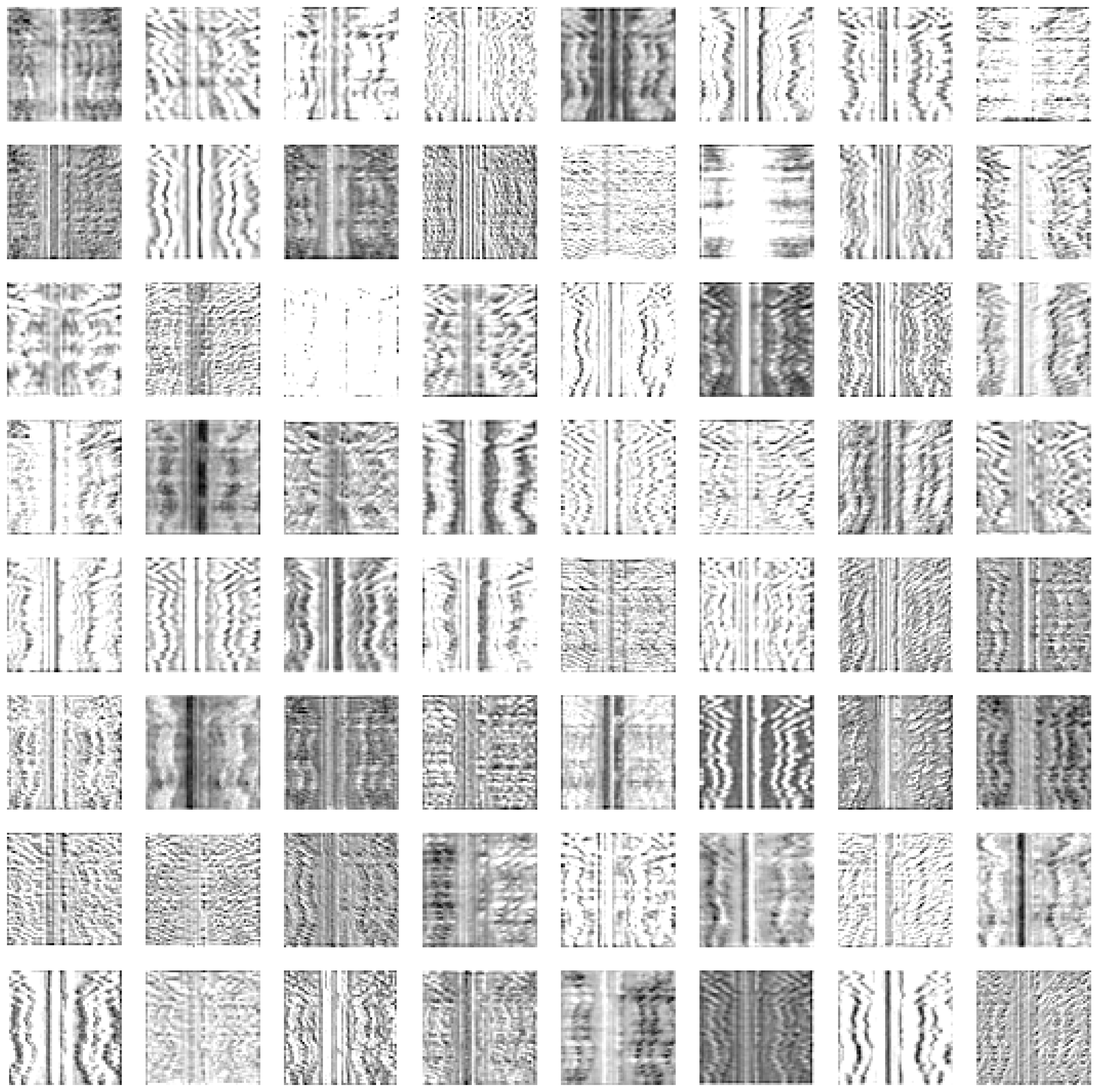}
\caption{Feature maps extracted from the carpet plot of LQTS patient (Fig. \ref{fig:carp1}) using ResNet18 model (first 3 convolutional layers).}
\label{fig:conv}
\end{figure}

To illustrate this potential, we applied a ResNet18 model \cite{He2016} pretrained on the ImageNet dataset \cite{Deng2009} to one of the carpet plots and visualized the resulting feature maps (Fig. \ref{fig:conv}). Even without training a dedicated classifier, the network was able to extract a variety of distinct patterns, underscoring that convolutional filters naturally respond to characteristic frequency-domain structures in the data, as explained by the convolution theorem. Some filters emphasize vertical morphology-related features, whereas others respond to longer-range temporal textures or periodic patterns. Our aim here is not to propose a specific classification model, but rather to demonstrate the feasibility of such integration. The clear differentiation of features across convolutional layers underscores the promise of combining carpet plot visualizations with convolutional networks or other AI-based methods to support automated analysis and anomaly detection.

Moreover, this representation aligns well with explainable AI (XAI) frameworks, enabling visual attribution of model decisions to specific regions of the carpet plot, thereby supporting transparent, interpretable classification or anomaly detection. In this way, our method serves as a unified platform that supports both expert-driven interpretation and automated, data-driven discovery. 

An additional advantage of this method is its ability to operate across multimodal signals, provided they are recorded simultaneously and properly synchronized. In such cases, the characteristic feature used for segmentation (e.g., R peaks) and the signal from which the segments are extracted and visualized can originate from different sources. For example, R peaks detected in a single ECG lead, such as lead I, can serve as temporal anchors for segmenting signals from other ECG channels, such as the precordial leads (V1--V6), enabling comparative visualization of spatial cardiac activity. This allows us to easily identify the epochs of transient decreases in conductance, which mainly affect QRS-duration dispersion. A separate paper on this subject is in preparation.

\begin{figure}
\centering
\begin{subfigure}[b]{0.32\textwidth}
         \centering
\includegraphics[width=\textwidth]{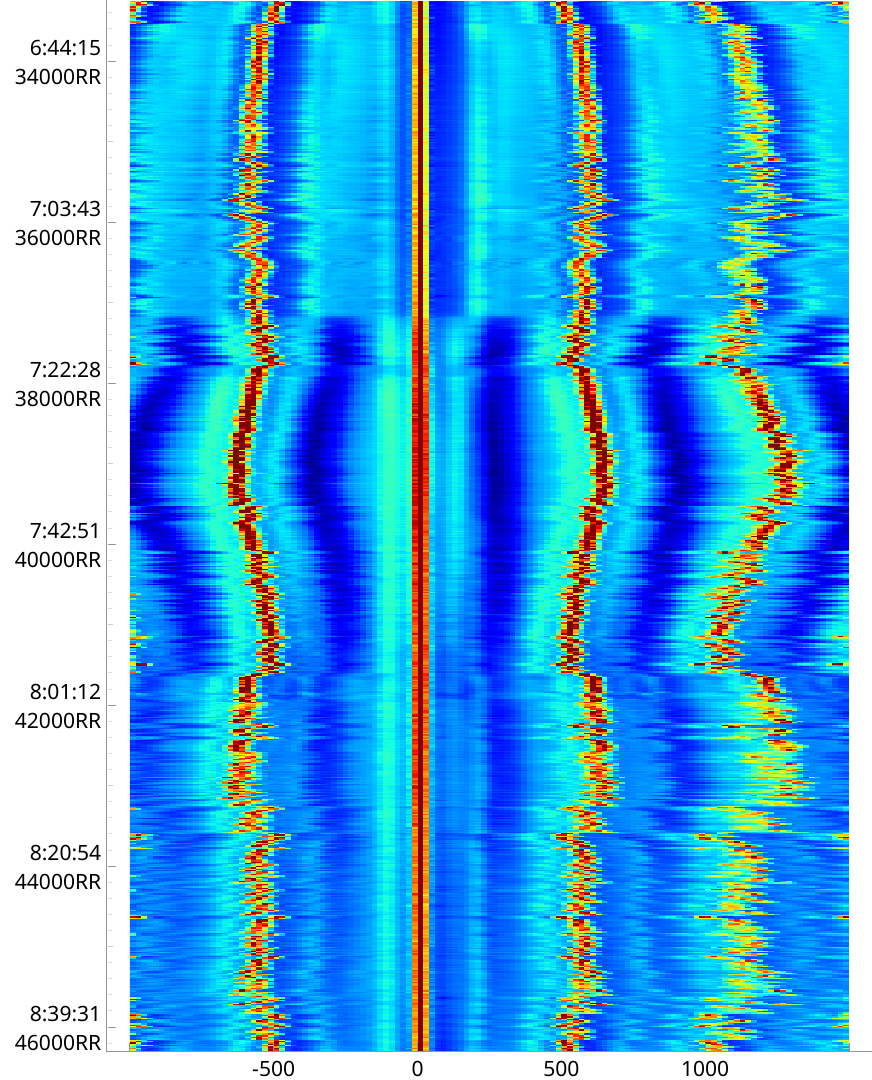}
\caption{ECG}
\label{fig:sync_ecg}
\end{subfigure}
\hfill
\begin{subfigure}[b]{0.32\textwidth}
         \centering
\includegraphics[width=\textwidth]{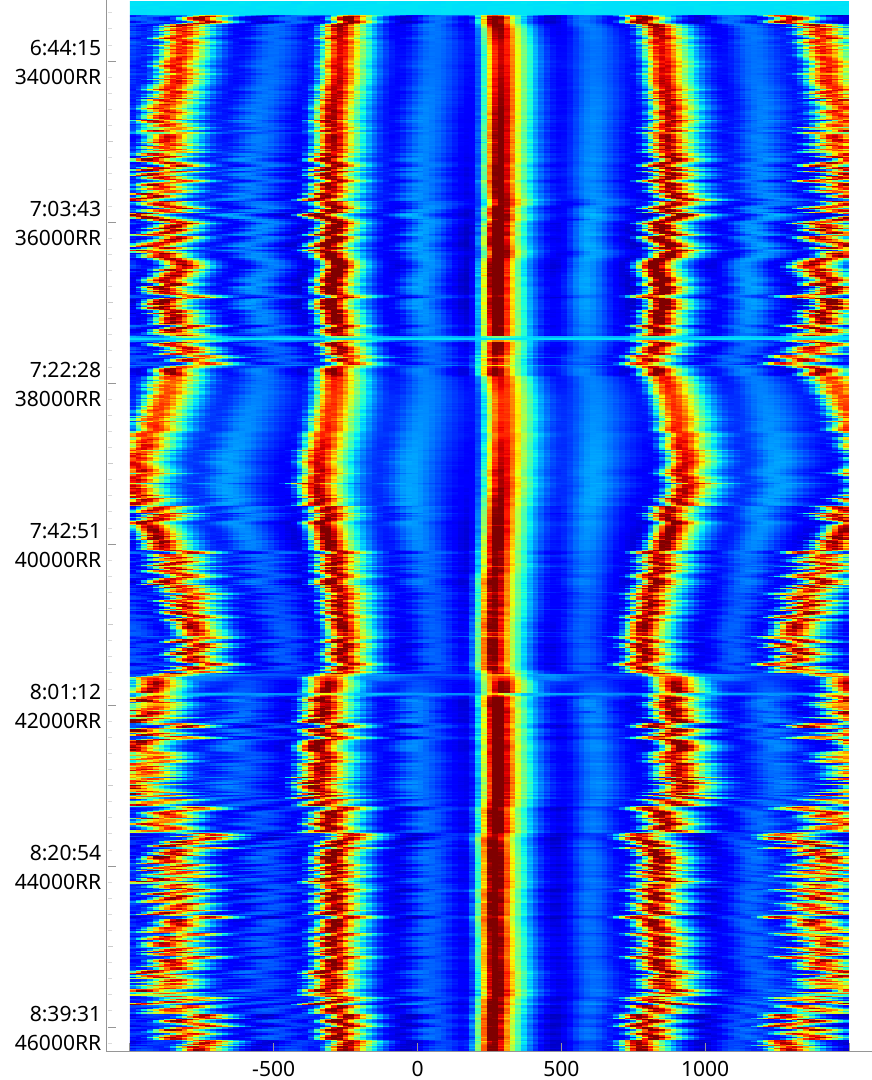}
\caption{ABP}
\label{fig:sync_abp}
\end{subfigure}
\hfill
\begin{subfigure}[b]{0.32\textwidth}
         \centering
\includegraphics[width=\textwidth]{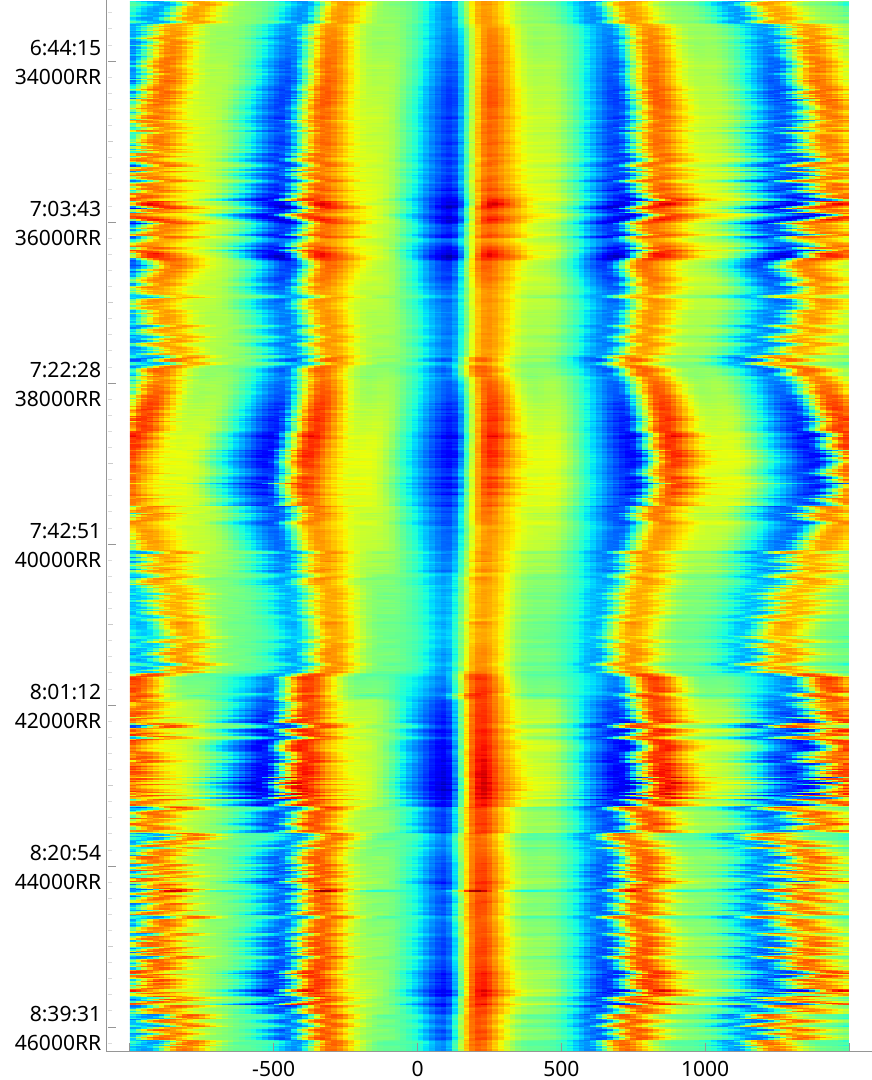}
\caption{ICP}
\label{fig:sync_icp}
\end{subfigure}
\hfill
\caption{Synchronized carpet plots of ECG, ABP, and ICP signals from the CHARIS database \cite{Kim2016} (record \#2).}
\label{fig:apb_icp}
\end{figure}

Finally, the same R peaks can be used to segment non-electrical physiological signals recorded in parallel, such as arterial blood pressure (ABP) or intracranial pressure (ICP). This approach enables the construction of carpet plots that reflect the dynamic coupling between cardiac rhythm and other physiological systems, providing a powerful tool for analyzing cardiovascular-cerebrovascular interactions, autonomic regulation, and more. By decoupling the reference feature from the target signal, the method supports flexible, synchronized exploration of complex biological processes (Fig. \ref{fig:apb_icp}). Although ECG, arterial blood pressure, and intracranial pressure exhibit substantially different waveform morphologies, synchronization using ECG-derived R peaks produces carpet plots with corresponding temporal organization. This enables direct visual comparison of transient changes across physiological systems and facilitates investigation of cardiorespiratory and cerebrovascular coupling. The application of the method to other biological signals, alignment using the PQ segment, and nonlinear color mapping are original contributions of this paper that distinguish it from \cite{Li2015} and \cite{Krasteva2025}, which we became aware of during the preparation of this manuscript.

\section{Robustness and limitations}

The proposed visualization method is robust to moderate physiological variability because each cardiac cycle is represented independently as a separate row of the carpet plot. Consequently, ectopic beats, premature atrial or ventricular contractions (PACs/PVCs), or transient rhythm disturbances remain localized within the visualization rather than being averaged together with neighboring beats. This preserves beat-to-beat variability and allows transient physiological events to be distinguished from persistent changes in rhythm or morphology.

The method also remains effective in the presence of irregular cardiac rhythms. For example, atrial fibrillation produces characteristic irregular vertical spacing, resulting from highly variable RR intervals and increased beat-to-beat morphological variability. Likewise, repetitive patterns such as ventricular bigeminy or trigeminy generate readily recognizable alternating structures. Rather than attempting to normalize or suppress these irregularities, the carpet representation preserves them, allowing both rhythm and morphology to be assessed simultaneously.

The practical robustness of the visualization depends not only on the quality of the recorded signal but also on the selected color-mapping strategy. When the full amplitude range of the recording is linearly mapped to the available color scale, a single high-amplitude artifact may dominate the dynamic range, substantially reducing the visibility of physiological features across the entire carpet plot. For this reason, we recommend percentile-based amplitude clipping, as described in the Methods section, which limits the influence of extreme outliers while preserving the contrast of the overwhelming majority of cardiac cycles. This simple preprocessing step substantially improves visualization quality for long-term recordings that frequently contain motion artifacts, electrode disturbances, or transient signal saturation.

Baseline wander is an important practical consideration because it directly affects the color mapping and may obscure subtle morphological changes. In our implementation, the ECG signal used for carpet plot generation is preprocessed using standard baseline correction consisting of power-line interference removal followed by high-pass filtering. For the human ECG recordings analyzed in this work, a 0.5 Hz high-pass filter provided satisfactory suppression of baseline wander while preserving waveform morphology. The optimal cutoff frequency, however, depends on the application. For example, in our work on small animals (rats) on respiratory function (paper in preparation), where both heart and respiratory rates are substantially higher than in humans, we found that a lower cutoff frequency (0.1 Hz) yielded superior results, whereas a 0.5 Hz cutoff introduced visible waveform distortions. Therefore, the preprocessing should be adapted to the physiological characteristics of the recorded signal.
    
The proposed visualization framework assumes that the fiducial points used for alignment (in this work, ECG R peaks) have been detected beforehand. R peak detection itself is outside the scope of the present study, as numerous dedicated algorithms are available for this purpose. In our implementation, R peaks were detected using the default detector provided by the NeuroKit2 library \cite{Makowski2021}, which is based on the steepness of the absolute ECG gradient and proved sufficiently robust for most of the datasets analyzed in this work. Before R peak detection, the ECG signal was preprocessed using a dedicated low-pass filter (50 Hz cutoff, Butterworth filter, order 5). This filtering was used exclusively for R peak detection; the carpet plots were generated from the original signal after standard preprocessing (power-line interference removal and 0.5 Hz high-pass filtering to correct for baseline wander).

In our experience, the default NeuroKit2 detector provided the most reliable performance across heterogeneous recordings. Classical Pan-Tompkins-type algorithms \cite{PanTompkins1985} and several of their derivatives produced a larger number of missed detections or false positives, particularly in recordings with pronounced T waves, such as long QT syndrome, where steep T wave upstrokes may be incorrectly identified as R peaks. For these challenging cases, the Elgendi algorithm \cite{Elgendi2013} and the wavelet-based method of Kalidas and Tamil \cite{Kalidas2017} demonstrated similarly robust performance.

The carpet-plot visualization is independent of the particular R peak detection algorithm. Missed or falsely detected R peaks affect only the corresponding rows in the carpet plot, resulting in local misalignment rather than degradation of the entire visualization. More advanced approaches, including machine learning-based methods \cite{HUQUE2019157, Chaoya2025} or algorithms that exploit nonlinear dynamics and chaos analysis \cite{GUPTA2019341}, may be employed when greater robustness is required. Consequently, the method can readily benefit from future improvements in R peak detection algorithms without modification of the visualization framework.

Although the method readily scales to multiple ECG leads through synchronized stacked carpet plots, displaying all 12 leads simultaneously may be impractical. In practice, interactive visualization or selection of clinically relevant leads provides a more readable alternative for long recordings.

\section{Conclusion}

In this work, we introduced a general visualization framework that transforms quasiperiodic physiological signals into two-dimensional carpet plots by aligning successive cycles to characteristic events and encoding signal amplitude using color mapping. Applied to ECG recordings, the proposed representation combines information on rhythm dynamics and beat morphology within a single visualization, enabling simultaneous assessment of temporal and morphological changes that are traditionally analyzed separately. The method was demonstrated using recordings from multiple publicly available databases containing normal rhythm, stress-test recordings, atrial fibrillation, premature ventricular contractions, atrioventricular block, and long QT syndrome, illustrating its applicability across a broad range of physiological conditions.

The presented examples show that carpet plots facilitate rapid identification of transient phenomena, including rhythm transitions, morphological alterations, and their temporal relationships. In particular, the visualization preserves beat-to-beat information while compressing long-duration recordings into a compact representation, making it practical for inspection of Holter ECGs and other extended physiological recordings. The method, therefore, complements conventional ECG visualization by providing an overview that is difficult to obtain from sequential waveform inspection alone.

Beyond visual interpretation, this representation provides a standardized image-like input suitable for computational analysis. As demonstrated using convolutional neural network feature extraction, carpet plots naturally integrate with modern machine learning approaches and may support automated detection, classification, and explainable analysis of physiological abnormalities. Furthermore, the same framework can be extended to synchronized multimodal recordings, enabling investigation of interactions between ECG and other physiological signals, such as arterial blood pressure or intracranial pressure.

This method is intended as a general-purpose visualization and analysis framework rather than a disease-specific diagnostic algorithm. Future work will focus on systematic clinical validation in larger patient populations, quantitative evaluation of diagnostic performance, extension to multilead ECG visualization, and development of dedicated machine learning models that fully exploit the rich spatiotemporal information contained in carpet plots.

\section{Acknowledgements}

Rafał Baranowski is acknowledged for his remarks, particularly for the PQ-based normalization idea. Mateusz Ozimek and Monika Petelczyc are thanked for the data. Natalia Dziura is thanked for testing.

\bibliographystyle{elsarticle-num} 
\bibliography{references}

@article{Hunter2007,
author = {John D. Hunter},
title = {Matplotlib: A {2D} Graphics Environment},
journal = {Computing in Science \& Engineering},
issue = {3},
volume = {9},
year = {2007},
pages = {90--95},
issn = {1521-9615}
}

@article{Goldberger2000,
author = {Ary L. Goldberger and Luis A. N. Amaral and Leon Glass and Jeffrey M. Hausdorff and Plamen Ch. Ivanov and Roger G. Mark and Joseph E. Mietus and George B. Moody and Chung-Kang Peng and H. Eugene Stanley},
title = {{PhysioBank}, {PhysioToolkit}, and {PhysioNet}},
journal = {Circulation},
issue = {23},
volume = {101},
year = {2000},
pages = {e215--e220},
issn = {0009-7322}
}

@article{Makowski2021,
author = {Dominique Makowski and Tam Pham and Zen J. Lau and Jan C. Brammer and François Lespinasse and Hung Pham and Christopher Schölzel and S. H. Annabel Chen},
title = {{NeuroKit2}: A {Python} toolbox for neurophysiological signal processing},
journal = {Behavior Research Methods},
issue = {4},
volume = {53},
year = {2021},
pages = {1689--1696},
issn = {1554-3528}
}

@misc{Kazemnejad2021,
  author = {Kazemnejad,  Arsalan and Gordany,  Peiman and Sameni,  Reza},
  title = {{EPHNOGRAM}: A Simultaneous Electrocardiogram and Phonocardiogram Database},
  publisher = {PhysioNet},
  year = {2021}
}

@article{Kim2016,
author = {Nam Kim and Alex Krasner and Colin Kosinski and Michael Wininger and Maria Qadri and Zachary Kappus and Shabbar Danish and William Craelius},
title = {Trending autoregulatory indices during treatment for traumatic brain injury},
journal = {Journal of Clinical Monitoring and Computing},
issue = {6},
volume = {30},
year = {2016},
pages = {821--831},
issn = {1387-1307}
}

@article{Acharya2006,
  author = {Rajendra Acharya,  U. and Paul Joseph,  K. and Kannathal,  N. and Lim,  Choo Min and Suri,  Jasjit S.},
  title = {Heart rate variability: a review},
  volume = {44},
  number = {12},
  journal = {Medical \& Biological Engineering \& Computing},
  publisher = {Springer Science and Business Media LLC},
  year = {2006},
  pages = {1031--1051}
}

@inproceedings{Andersen2007,
   author = {M.P. Andersen and J.Q. Xue and C. Graff and T.B. Hardahl and E. Toft and J.K. Kanters and M. Christiansen and H.K. Jensen and J.J. Struijk},
   isbn = {978-1-4244-2533-4},
   booktitle = {2007 Computers in Cardiology},
   pages = {341--344},
   publisher = {IEEE},
   title = {A robust method for quantification of {IKr}-related {T-wave} morphology abnormalities},
   year = {2007}
}

@article{Malik2002,
author = {M Malik},
title = {Relation between {QT} and {RR} intervals is highly individual among healthy subjects: implications for heart rate correction of the {QT} interval},
journal = {Heart},
issue = {3},
volume = {87},
year = {2002},
pages = {220--228},
issn = {00070769}
}

@article{Okin2002,
author = {Peter M. Okin and Richard B. Devereux and Richard R. Fabsitz and Elisa T. Lee and James M. Galloway and Barbara V. Howard},
title = {Principal Component Analysis of the {T Wave} and Prediction of Cardiovascular Mortality in {American Indians}},
journal = {Circulation},
issue = {6},
volume = {105},
year = {2002},
pages = {714--719},
issn = {0009-7322}
}

@inproceedings{Andrzejewska2022,
   author = {Malgorzata Andrzejewska and Mateusz Ozimek and Karolina Rams and Teodor Buchner},
   isbn = {978-1-6654-8512-8},
   booktitle = {2022 12th Conference of the European Study Group on Cardiovascular Oscillations (ESGCO)},
   pages = {1--2},
   publisher = {IEEE},
   title = {Asymmetry of {RR} intervals and {ECG} amplitudes in {LQTS} patients},
   year = {2022}
}

@article{Peczalski2024,
author = {Kazimierz Pęczalski and Judyta Sobiech and Teodor Buchner and Thomas Kornack and Elizabeth Foley and Daniel Janczak and Małgorzata Jakubowska and David Newby and Nancy Ford and Maryla Zajdel},
title = {Synchronous recording of magnetocardiographic and electrocardiographic signals},
journal = {Scientific Reports},
issue = {1},
volume = {14},
year = {2024},
pages = {4098},
issn = {20452322},
pmid = {38374368},
publisher = {Nature Research}
}

@article{Boyett2000,
author = {M Boyett},
title = {The sinoatrial node, a heterogeneous pacemaker structure},
journal = {Cardiovascular Research},
issue = {4},
volume = {47},
year = {2000},
pages = {658--687},
issn = {00086363}
}

@book{Glass1991,
   city = {New York, NY},
   editor = {Leon Glass and Peter Hunter and Andrew McCulloch},
   isbn = {978-1-4612-7803-0},
   publisher = {Springer New York},
   title = {Theory of Heart},
   year = {1991}
}

@inproceedings{GualsaquiMiranda2016,
  author={Gualsaquí Miranda, Marco V. and Vizcaíno Espinosa, Iván P. and Flores Calero, Marco J.},
  booktitle={2016 IEEE Ecuador Technical Chapters Meeting (ETCM)}, 
  title={{ECG} signal features extraction}, 
  year={2016},
  volume={},
  number={},
  pages={1--6},
}

@inproceedings{Li2011,
   author = {Fei Li and Jie Zhao and Hui Lin Jia and Chun Yun Zhang and Xiao Lei Zhu},
   issn = {18780296},
   booktitle = {Procedia Environmental Sciences},
   pages = {575--581},
   publisher = {Elsevier B.V.},
   title = {Poincaré mapping: A potential method for detection of {T-wave} alternans},
   volume = {8},
   year = {2011}
}

@inproceedings{Maniewski1996,
  author={Maniewski, R. and Lewandowski, P. and Nowinska, M. and Mroczka, T.},
  booktitle={Proceedings of 18th Annual International Conference of the IEEE Engineering in Medicine and Biology Society}, 
  title={Time-frequency methods for high-resolution {ECG} analysis}, 
  year={1996},
  volume={3},
  number={},
  pages={1266--1267},
}

@inproceedings{Couderc2010,
   author = {Jean Philippe Couderc},
   isbn = {9781424441235},
   booktitle = {2010 Annual International Conference of the IEEE Engineering in Medicine and Biology Society, EMBC'10},
   pages = {6252--6255},
   pmid = {21097349},
   title = {The Telemetric and {Holter} {ECG} Warehouse initiative ({THEW}): A data repository for the design, implementation and validation of {ECG}-related technologies},
   year = {2010}
}

@article{Moody1983,
author = {George Moody and Roger Mark},
title = {A new method for detecting atrial fibrillation using {R-R} intervals},
journal = {Computers in Cardiology},
issue = {1983},
year = {1983},
pages = {227--230}
}

@inproceedings{He2016,
   author = {Kaiming He and Xiangyu Zhang and Shaoqing Ren and Jian Sun},
   isbn = {978-1-4673-8851-1},
   booktitle = {2016 IEEE Conference on Computer Vision and Pattern Recognition (CVPR)},
   pages = {770--778},
   publisher = {IEEE},
   title = {Deep Residual Learning for Image Recognition},
   year = {2016}
}

@inproceedings{Deng2009,
   author = {Jia Deng and Wei Dong and Richard Socher and Li-Jia Li and Kai Li and Li Fei-Fei},
   isbn = {978-1-4244-3992-8},
   booktitle = {2009 IEEE Conference on Computer Vision and Pattern Recognition},
   pages = {248--255},
   publisher = {IEEE},
   title = {{ImageNet}: A large-scale hierarchical image database},
   year = {2009}
}

@article{Brubaker2011,
author = {Brubaker, Peter H. and Kitzman, Dalane W.},
title = {{Chronotropic Incompetence}},
journal = {Circulation},
volume = {123},
year = {2011},
pages = {1010--1020},
issn = {0009-7322},
number = {9}
}

@article{Kuklik2009,
author = {Kuklik, P. and Szumowski, L. and {\.{Z}}ebrowski, J. J. and Sanders, P.},
title = {{Integration of the data from electroanatomical mapping system and CT imaging modality}},
journal = {The International Journal of Cardiovascular Imaging},
volume = {25},
year = {2009},
pages = {425--432},
issn = {1569-5794},
number = {4}
}

@article{Wagner2009,
author = {Wagner, Galen S. and MacFarlane, Peter and Wellens, Hein and Josephson, Mark and Gorgels, Anton and Mirvis, David M. and Pahlm, Olle and Surawicz, Borys and Kligfield, Paul and Childers, Rory and Gettes, Leonard S.},
title = {{AHA/ACCF/HRS recommendations for the standardization and interpretation of the electrocardiogram}},
journal = {Circulation},
volume = {119},
year = {2009},
pages = {262--270},
issn = {00097322},
number = {10},
pmid = {19228819}
}

@article{Junttila2012,
author = {Junttila, M. J. and Sager, S. J. and Tikkanen, J. T. and Anttonen, O. and Huikuri, H. V. and Myerburg, R. J.},
title = {{Clinical significance of variants of J-points and J-waves: early repolarization patterns and risk}},
journal = {European Heart Journal},
volume = {33},
year = {2012},
pages = {2639--2643},
issn = {0195-668X},
number = {21}
}

@article{DeAndrade2022,
author = {de Andrade, Antonio Thomaz and Barbosa-Barros, Raimundo and Nikus, Kjell and Raimundo, Rodrigo D. and de Abreu, Luiz C. and Sacilotto, Luciana and Darriuex, Francisco C.C. and Yanowitz, Frank G. and Brugada, Pedro and P{\'{e}}rez-Riera, Andr{\'{e}}s Ricardo},
title = {{Transient ascending ST-segment depression and widening of the S wave in 3-channel {Holter} monitoring}},
journal = {Annals of Noninvasive Electrocardiology},
volume = {27},
year = {2022},
pages = {1--6},
issn = {1542474X},
number = {2},
pmid = {34882891}
}

@article{Bjerregaard2003,
author = {Bjerregaard, Preben and El-Shafei, Amr and Kotar, Susan L. and Labovitz, Arthur J.},
title = {{ST segment analysis by {Holter} Monitoring: Methodological considerations}},
journal = {Annals of Noninvasive Electrocardiology},
volume = {8},
year = {2003},
pages = {200--207},
issn = {1082720X},
number = {3},
pmid = {14510654}
}

@article{Narayan2006,
author = {Narayan, Sanjiv M.},
title = {{T-wave alternans and the susceptibility to ventricular arrhythmias}},
journal = {Journal of the American College of Cardiology},
volume = {47},
year = {2006},
pages = {269--281},
issn = {07351097},
number = {2},
pmid = {16412847},
publisher = {Elsevier Masson SAS}
}

@article{Cutler2009,
author = {Cutler, Michael J and Rosenbaum, David S},
title = {{Explaining the clinical manifestations of T wave alternans in patients at risk for sudden cardiac death}},
journal = {Heart Rhythm},
volume = {6},
year = {2009},
pages = {S22--8},
issn = {1556-3871},
number = {3 Suppl},
pmid = {19168395}
}

@article{Aro2016,
author = {Aro, Aapo L and Kentt{\"{a}}, Tuomas V and Huikuri, Heikki V},
issn = {2050-3369},
journal = {Arrhythmia \& Electrophysiology Review},
number = {1},
pages = {37--40},
title = {{Microvolt T-wave Alternans: Where Are We Now?}},
volume = {5},
year = {2016}
}

@article{Gabr2023,
author = {Gabr, Mohamed and Mangeshkar, Shaunak and Cerna, Luis and Rahgozar, Kusha and Al-Taei, Mustafa H and Grushko, Michael},
issn = {0009-7322},
journal = {Circulation},
number = {Suppl\_1},
title = {{Abstract 11663: T Wave Alternans: A Clue in Plain Sight}},
volume = {148},
year = {2023}
}

@article{Starobin2009,
author = {Starobin, Joseph M. and Danford, Christopher P. and Varadarajan, Vivek and Starobin, Andrei J. and Polotski, Vladimir N.},
issn = {17534631},
journal = {Nonlinear Biomedical Physics},
pages = {1--7},
title = {{Critical scale of propagation influences dynamics of waves in a model of excitable medium}},
volume = {3},
year = {2009}
}

@article{Krogh-Madsen2007,
author = {Krogh-Madsen, Trine and Christini, David J.},
issn = {00063495},
journal = {Biophysical Journal},
number = {4},
pages = {1138--1149},
pmid = {17114216},
publisher = {Elsevier},
title = {{Action potential duration dispersion and alternans in simulated heterogeneous cardiac tissue with a structural barrier}},
volume = {92},
year = {2007}
}

@article{Li2015,
   author = {Duan Li and Fangyun Tian and Santiago Rengifo and Gang Xu and Michael M. Wang and Jimo Borjigin},
   issue = {5},
   journal = {Journal of Integrative Cardiology},
   publisher = {Open Access Text Pvt, Ltd.},
   title = {Electrocardiomatrix: A new method for beat-by-beat visualization and inspection of cardiac signals},
   volume = {1},
   year = {2015},
   pages = {133}
}

@article{Krasteva2025,
   author = {Vessela Krasteva and Todor Stoyanov and Stefan Naydenov and Ramun Schmid and Irena Jekova},
   issn = {20754418},
   issue = {7},
   journal = {Diagnostics},
   publisher = {Multidisciplinary Digital Publishing Institute (MDPI)},
   title = {Detection of Atrial Fibrillation in {Holter} {ECG} Recordings by {ECHOView} Images: A Deep Transfer Learning Study},
   volume = {15},
   year = {2025},
   pages = {865}
}

@article{Turcott1994,
  title = {A nonstationary {Poisson} point process describes the sequence of action potentials over long time scales in lateral-superior-olive auditory neurons},
  volume = {70},
  ISSN = {1432-0770},
  number = {3},
  journal = {Biological Cybernetics},
  publisher = {Springer Science and Business Media LLC},
  author = {Turcott,  Robert G. and Lowen,  Steven B. and Li,  Eric and Johnson,  Don H. and Tsuchitani,  Chiyeko and Teich,  Malvin C.},
  year = {1994},
  pages = {209--217}
}

@article{taskforce1996hrv,
  title        = {Heart rate variability: Standards of measurement, physiological interpretation, and clinical use},
  author       = {{Task Force of the European Society of Cardiology and the North American Society of Pacing and Electrophysiology}},
  journal      = {Circulation},
  volume       = {93},
  number       = {5},
  pages        = {1043--1065},
  year         = {1996},
}

@ARTICLE{Hoshide2022-hw,
  title     = "Pulse transit time-estimated blood pressure: a comparison of
               beat-to-beat and intermittent measurement",
  author    = "Hoshide, Satoshi and Yoshihisa, Akiomi and Tsuchida, Fumihiro
               and Mizuno, Hiroyuki and Teragawa, Hiroki and Kasai, Takatoshi
               and Koito, Hitoshi and Ando, Shin-Ichi and Watanabe, Yoshihiko
               and Takeishi, Yasuchika and Kario, Kazuomi",
  journal   = "Hypertens. Res.",
  volume    =  45,
  number    =  6,
  pages     = "1001--1007",
  year      =  2022,
}

@article{Efimov,
author = {Igor R. Efimov  and Vladimir P. Nikolski  and Guy Salama },
title = {Optical Imaging of the Heart},
journal = {Circulation Research},
volume = {95},
number = {1},
pages = {21--33},
year = {2004},
}

@ARTICLE{Kessler1992,
  title     = "{ST-alternans} alternans",
  author    = "Kessler, K M and Bauerlein, E J and Schob, A and de Marchena, E
               and Myerburg, R J",
  journal   = "Am. J. Cardiol.",
  publisher = "Elsevier BV",
  volume    =  70,
  number    =  13,
  pages     = "1224--1225",
  year      =  1992,
  language  = "en",
}

@article{Baranowski2002,
  title = {Assessment of the {RR} versus {QT} relation by a new symbolic dynamics method},
  volume = {35},
  number = {2},
  journal = {Journal of Electrocardiology},
  publisher = {Elsevier BV},
  author = {Baranowski,  Rafa{\l} and {\.Z}ebrowski,  Jan J.},
  year = {2002},
  pages = {95--103},
}

@article{Baumert2015,
  title = {Joint symbolic dynamics for the assessment of cardiovascular and cardiorespiratory interactions},
  volume = {373},
  number = {2034},
  journal = {Philosophical Transactions of the Royal Society A: Mathematical,  Physical and Engineering Sciences},
  author = {Baumert,  Mathias and Javorka,  Michal and Kabir,  Muammar},
  year = {2015},
  pages = {20140097},
}

@article{Hastings2000,
  title = {Alternans and the onset of ventricular fibrillation},
  volume = {62},
  ISSN = {1095-3787},
  number = {3},
  journal = {Physical Review E},
  publisher = {American Physical Society (APS)},
  author = {Hastings,  Harold M. and Fenton,  Flavio H. and Evans,  Steven J. and Hotomaroglu,  Omer and Geetha,  Jagannathan and Gittelson,  Ken and Nilson,  John and Garfinkel,  Alan},
  year = {2000},
  pages = {4043--4048}
}

@misc{github,
    author = {Gradowski, Tomasz and Waląg, Damian and Domański, Tomir and Buchner, Teodor},
  title = {{ECG} Carpets},
  year = {2025},
  publisher = {GitHub},
  journal = {GitHub repository},
  howpublished = {\url{https://github.com/tomgrad/carpets}}
}

@article{PENG2004199,
title = {Application of the wavelet transform in machine condition monitoring and fault diagnostics: a review with bibliography},
journal = {Mechanical Systems and Signal Processing},
volume = {18},
number = {2},
pages = {199--221},
year = {2004},
issn = {0888-3270},
author = {Z.K. Peng and F.L. Chu}
}

@article{Stiles2004,
  title = {Wavelet‐Based Analysis of Heart‐Rate‐Dependent {ECG} Features},
  volume = {9},
  ISSN = {1542-474X},
  number = {4},
  journal = {Annals of Noninvasive Electrocardiology},
  publisher = {Wiley},
  author = {Stiles,  Martin K. and Clifton,  David and Grubb,  Neil R. and Watson,  James N. and Addison,  Paul S.},
  year = {2004},
  pages = {316--322}
}

@INPROCEEDINGS{Mewett1999,
  author={Mewett, D.T. and Reynolds, K.J. and Nazeran, H.},
  booktitle={Proceedings of the First Joint BMES/EMBS Conference. 1999 IEEE Engineering in Medicine and Biology 21st Annual Conference and the 1999 Annual Fall Meeting of the Biomedical Engineering Society}, 
  title={Recurrence plot features of {ECG} signals}, 
  year={1999},
  volume={2},
  number={},
  pages={913}}

@article{GUPTA2019341,
title = {R-Peak Detection Using Chaos Analysis in Standard and Real Time {ECG} Databases},
journal = {IRBM},
volume = {40},
number = {6},
pages = {341--354},
year = {2019},
issn = {1959-0318},
author = {V. Gupta and M. Mittal and V. Mittal},
}

@article{HUQUE2019157,
title = {{HMM}-based Supervised Machine Learning Framework for the Detection of {fECG} {R-R} Peak Locations},
journal = {IRBM},
volume = {40},
number = {3},
pages = {157--166},
year = {2019},
issn = {1959-0318},
author = {A.S.A. Huque and K.I. Ahmed and M.A. Mukit and R. Mostafa},
}

@article{Chaoya2025,
  title = {Robust R-peak detection in noisy {ECG} using deep residual {U-Net} for enhanced cardiac rhythm analysis},
  volume = {104},
  ISSN = {1536-5964},
  number = {40},
  journal = {Medicine},
  publisher = {Ovid Technologies (Wolters Kluwer Health)},
  author = {Chaoya,  Wang and Chun,  Pan and Chao,  Meng},
  year = {2025},
  pages = {e44968}
}

@article{WILDOWICZ2025199,
title = {Physically motivated projection of the electrocardiogram—A feasibility study},
journal = {Biocybernetics and Biomedical Engineering},
volume = {45},
number = {2},
pages = {199-211},
year = {2025},
issn = {0208-5216},
author = {Sebastian Wildowicz and Tomasz Gradowski and Paulina Figura and Igor Olczak and Judyta Sobiech and Teodor Buchner},
}

@ARTICLE{PanTompkins1985,
  author={Pan, Jiapu and Tompkins, Willis J.},
  journal={IEEE Transactions on Biomedical Engineering}, 
  title={A Real-Time {QRS} Detection Algorithm}, 
  year={1985},
  volume={BME-32},
  number={3},
  pages={230-236}}

@INPROCEEDINGS{Kalidas2017,
  author={Kalidas, Vignesh and Tamil, Lakshman},
  booktitle={2017 IEEE 17th International Conference on Bioinformatics and Bioengineering (BIBE)}, 
  title={Real-time {QRS} detector using Stationary Wavelet Transform for Automated {ECG} Analysis}, 
  year={2017},
  volume={},
  number={},
  pages={457-461}
}

@article{Elgendi2013,
    author = {Elgendi, Mohamed},
    journal = {PLOS ONE},
    publisher = {Public Library of Science},
    title = {Fast {QRS} Detection with an Optimized Knowledge-Based Method: Evaluation on 11 Standard {ECG} Databases},
    year = {2013},
    month = {09},
    volume = {8},
    pages = {1-18},
    number = {9}
}

@article{SHARKEY2026102671,
title = {Performance of Artificial Intelligence-Powered {ECG} Analysis in Suspected {ST}-Segment Elevation Myocardial Infarction},
journal = {JACC: Advances},
volume = {5},
number = {4},
pages = {102671},
year = {2026},
issn = {2772-963X},
author = {Scott W. Sharkey and Robert Herman and Dawn R. Witt and Frank Aguirre and Mehmet Yildiz and David M. Larson and Avinash Murthy and Heather S. Rohm and Stephen W. Smith and Will Belzer and Jenny Chambers and Ellen Cravero and Seth Bergstedt and Greg Kerola and David Farmer and Andrew Willett and H. Pendell Meyers and Julia Harris and Christopher VanHove and Timothy D. Henry},
}

@article{Kim2026,
    author = {Kim, Y R and Lee, K H and Yoon, N S and Park, H W and Yang, H J},
    title = {Artificial intelligence-assisted {QT} interval measurement in 12-lead {ECGs}},
    journal = {European Heart Journal - Digital Health},
    volume = {7},
    number = {Supplement\_1},
    pages = {ztaf143.129},
    year = {2026},
    month = {01},
    issn = {2634-3916},
}

\end{document}